\documentclass{iopjournal}
\usepackage{amsmath}
\usepackage{amssymb}
\usepackage{xcolor}
\usepackage{array}
\usepackage{booktabs}
\usepackage{tensor}
\usepackage{color}
\usepackage{float} % for [H]
\usepackage{subcaption}
\usepackage{comment}
\usepackage{cite}
\usepackage{ragged2e}

\newcommand{\bs}{\begin{split}}
\newcommand{\es}{\end{split}}

\begin{document}

%\articletype{Paper} %	 e.g. Paper, Letter, Topical Review...
\begin{justify}
\title{Energy Extraction via Magnetic Reconnection from a Rotating Dyonic Black Hole in $\mathcal{N} = 2, \ U(1)^2$ Gauged Supergravity}

\author{Raid M Suleiman$^{1,*}$\orcid{0000-0003-1937-1189}, Shanshan Rodriguez$^{1,2}$\orcid{0000-0003-2944-0424}, Dominic O. Chang$^{1,3}$\orcid{0000-0001-9939-5257}, and  Leo Rodriguez$^{2}$\orcid{0000-0003-4800-2921}}

\affil{$^1$Center for Astrophysics $|$ Harvard \& Smithsonian, Cambridge, US.}

\affil{$^2$Physics Department, Grinnell College, Grinnell, US.}

\affil{$^3$Black Hole Initiative, Harvard University, Cambridge, US.}

\affil{$^*$Author to whom any correspondence should be addressed.}

\email{rsuleiman@cfa.harvard.edu}


\begin{abstract}
\begin{justify}
We study energy extraction via magnetic reconnection from a rotating dyonic black hole in four-dimensional $\mathcal{N}=2$, $U(1)^2$ gauged supergravity. Using the Comisso–Asenjo mechanism in the ZAMO frame, we derive the asymptotic hydrodynamic energy per unit enthalpy $\epsilon^{\infty}_\pm$ and determine when reconnection outflows attained negative energy at infinity. By varying the spin $a$, electric and magnetic charges, NUT parameter $N_g$, and gauge coupling $g$, we compute the cutoff magnetization $\sigma_0^{\rm cutoff}$ and map the region of parameter space that admits $\epsilon^{\infty}_-<0$.
We find that $\sigma_0^{\rm cutoff}$ and the very existence of Comisso–Asenjo extraction are tightly controlled by $g$ and the dyonic charges: increasing $g$ or pushing the charges toward extremality raises $\sigma_0^{\rm cutoff}$ and shrinks the CA‑active part of the ergoregion. Unlike Kerr, the spin enters through the normalization factor $\Xi$, and the quartic horizon function $\Delta_g$, so geometric effects from the AdS/NUT deformation dominate the usual frame‑dragging enhancement. As a result, the extracted power and efficiency are non-monotonic in $a$ and peak at intermediate spin ($a\sim0.8$); near-extremal rotation is not required for high efficiency, provided $g$ is small and $Q$ is moderate. Efficient extraction further demands extreme magnetization and nearly radial outflows, confining the active reconnection layer to a thin shell, just outside the horizon.

\end{justify}
\end{abstract}

\section{Introduction}\label{sec:1}
Black holes are among the most striking predictions of general relativity and now occupy a central place in high‑energy astrophysics. The first exact solution, Schwarzschild’s static, spherically symmetric spacetime \cite{schwarzschild1916}, is too idealized for most astrophysical situations, where collapse and accretion naturally produce rotating objects. Allowing for spin leads to the Kerr metric \cite{Kerr1963}, in which a black hole is completely characterized by its mass and angular momentum. The existence of such Kerr‑like compact objects is supported independently by gravitational wave observations of binary mergers \cite{Abbott2016,Abbott2017} and horizon‑scale images of the supermassive black holes in M87* and SgrA* \cite{EHT2019,Akiyama2022}.

Kerr black holes are widely believed to power some of the most energetic phenomena in the universe. At the stellar-mass scale, accreting Kerr black holes drive X-ray binaries and gamma-ray bursts, where accretion flows and relativistic jets explain the observed high energy emission \cite{Remillard2006,McClintock2014,Corral2016,Woosley2006,Cano2017,Reynolds2021}. At supermassive scales, quasars and active galactic nuclei are interpreted as powered by accretion onto Kerr black holes in galactic nuclei \cite{Kormendy2013,Heckman2014,Netzer2015}.
Unlike the Schwarzschild spacetime, the Kerr spacetime permits the extraction of rotational energy and angular momentum via frame dragging. 

Unlike the Schwarzschild solution, the Kerr spacetime allows one, at least in principle, to tap into the rotational energy of the black hole through frame dragging. Penrose \cite{Penrose1971} first showed that this could be done via particle orbits in the ergosphere that carry negative energy (as measured at infinity), so that their capture spins the hole down. Variants of this idea have been explored in more elaborate settings, including magnetized plasmas and supergravity backgrounds. However, as Wald emphasized \cite{Wald1974}, the original Penrose process is probably irrelevant for real astrophysical sources: it requires such finely tuned orbits that the net power output is vanishingly small.

In realistic astrophysical environments, efficient energy extraction is instead expected to occur via plasma‑mediated processes, in which collective electromagnetic fields naturally generate negative energy states at macroscopic rates. This perspective has motivated several magnetohydrodynamic generalizations of the Penrose mechanism. Prominent among these is the Blandford–Znajek process \cite{Blandford1977ds, Meringolo2025}, in which large‑scale magnetic fields threading the horizon extract rotational energy electromagnetically, transfer it to the surrounding plasma, and power relativistic jets. Related frameworks include the MHD Penrose process \cite{Takahashi1990, Ruffini1975}, where magnetized plasma flows enable negative‑energy inflow, and ergospheric magnetic reconnection scenarios \cite{Koide2008}, in which dissipation and changes in field‑line topology facilitate energy release. Taken together, these mechanisms offer a plausible way to tap into the black hole's rotational energy in realistic astrophysical settings.

Recently, Comisso and Asenjo \cite{Comisso2021} showed that fast magnetic reconnection in the ergosphere can extract rotational energy at astrophysically relevant rates by converting magnetic energy in a highly conducting plasma, rather than relying on finely tuned particle orbits as in the classical Penrose process. In fast reconnection, antiparallel magnetic field lines are rapidly rearranged in a thin current sheet, converting magnetic energy into bulk kinetic, thermal, and non‑thermal particle energy on macroscopic scales.
In the Comisso–Asenjo scenario, reconnection layers form inside the ergosphere with oppositely directed outflows. Plasma expelled along one direction is accelerated in the sense of the black hole spin and can escape to infinity, while plasma expelled in the opposite direction can be driven onto orbits with negative energy as seen by a distant observer and fall into the horizon. When this happens, the outgoing component carries away more energy than the incoming plasma supplied, and the difference is drawn from the black hole’s rotational energy reservoir. Negative energy states are thus produced collectively by reconnection‑driven plasma dynamics, rather than by carefully arranged individual particle collisions.

In recent years, the Comisso–Asenjo mechanism has been used as a probe of deviations from the Kerr geometry in a variety of rotating black hole backgrounds. Closest to the standard Kerr–Newman family are studies in modified Kerr–Newman spacetimes, both in alternative theories of gravity \cite{Shaymatov2024a} and in perfect fluid dark matter halos \cite{Rodriguez2025}, as well as analyses in asymptotically AdS geometries such as Kerr–de Sitter, Kerr–Sen–AdS$_4$, and higher‑dimensional Myers–Perry black holes \cite{Wang2022,Shaymatov2024b,Zeng2025}. These studies trace how a cosmological constant, additional gauge fields, or extra dimensions affect the efficiency and parameter space of CA extraction.

Beyond these Kerr–Newman-like and AdS cases, the mechanism has been explored in several non-Kerr and phenomenological metrics. Examples include spinning braneworld black holes with tidal charge \cite{Wei2008}, Kerr–MOG solutions in scalar–tensor–vector gravity \cite{Khodadi2023}, Lorentz‑violating Kerr–Sen and Kiselev spacetimes \cite{Carleo2022}, and various parametrized or regular and hairy rotating black holes \cite{Liu2022,Li2023a,Li2023b,Zhang2024a,Long2025}. More exotic or topologically non‑trivial geometries have also been considered, such as spinning wormholes \cite{Ye2023}, Kerr Taub–NUT and Kerr Bertotti–Robinson backgrounds \cite{Cheng2025,Zeng2025a}, and accelerating Kerr black holes \cite{Wang-Zeng2025}.

In parallel with these background‑oriented studies, a number of works focus on the plasma physics of the process itself. They examined reconnection in plunging and circular orbit regions \cite{Chen2024,Shen2024}, the impact of intrinsic black hole magnetization \cite{Zhang2024b}, and the role of field orientation angles and reconnection parameters \cite{Shen2024,Shen2025} in shaping the efficiency and variability of CA extraction.

Motivated by these previous works, we are interested in investigating the energy extraction mechanism in supergravity (SUGRA) backgrounds when multiple gauge charges and non-trivial scalar profiles reshape both the ergoregion and horizon structure, thus leading to new features during its interaction with its surrounding plasmas. The basic family of charged rotating black holes in four-dimensional gauged and ungauged supergravity with four independent electromagnetic charges was constructed by \cite{Chong2004}, who exploited the global $O(4,4)$ symmetry of dimensionally reduced three-dimensional theory. In the ungauged case, acting on the uncharged Kerr metric with an $O(1,1)^4 \subset O(4,4)$ transformation produces a rotating black hole carrying four independent electromagnetic charges, which can be viewed as a solution of $\mathcal{N}=2$ supergravity coupled to three abelian vector multiplets. Previous studies indicate that such extra charges and matter fields can strongly affect orbital structure. For example, \cite{Shaymatov2022} found that charged spinning particles around a dyonic black hole in a dark matter halo experience large shifts of the ISCO radius as the black hole charges and particle spin are varied.

Rotating supergravity black holes, such as the STU family, dilaton–axion solutions, and minimal gauged supergravities in four and five dimensions, have been extensively studied in terms of their thermodynamics, supersymmetry, and superradiant stability \cite{Cvetic2005,Lu2013,Aliev2009,Cvetic2020,Mukherjee2016,Gnecchi2014,Sharma2021,Cvetic2018}. However, the role of these supergravity-specific structures in plasma-mediated energy extraction remains largely unexplored. In particular, it is not yet clear how the combined effects of multiple gauge charges, the gauge coupling $g$, and the associated scalar fields reshape the horizon and ergoregion so as to support the negative energy plasma states required for magnetic reconnection.

A natural testing ground is provided by the rotating dyonic black hole of four‑dimensional $\mathcal{N}=2$, $U(1)^2$ gauged supergravity constructed by \cite{Chong2004}, which carries two independent electric/magnetic charge pairs and non‑trivial scalar profiles. The impact of the gauge coupling on classical Penrose extraction in this background has been analyzed by Rudra \ \cite{Rudra2020}, who showed that, under suitable charge and coupling constraints, the maximal Penrose efficiency can approach $\sim 60.75\%$ of the black hole mass, nearly twice the extremal Kerr value. This suggests that supergravity charges and gauging can substantially enhance or reshape the available energy reservoir. In this work we extend this line of inquiry to the Comisso–Asenjo mechanism, we study fast magnetic reconnection, in the sense of \cite{Comisso2021}, in the rotating dyonic $\mathcal{N}=2$, $U(1)^2$ gauged supergravity black hole, following how the spin $a$, electric charge $Q$, magnetic charge $v$, NUT charge $N_g$, and gauge coupling $g$ affect the horizon and ergoregion geometry, the parameter space that admits negative energy plasma states, and the resulting CA power and efficiency. This will allow a direct comparison with the Kerr and Kerr–Newman cases and a quantitative assessment of the impact of supergravity specific structure on reconnection‑driven energy extraction.

The remainder of this paper is organized as follows: In Section \ref{sec:2}, we introduce the rotating dyonic black hole in $\mathcal{N} = 2, \ U(1)^2$ gauged supergravity and show the impact of the black hole spin $a$, electric charge $Q$, and gauge coupling $g$ on the event horizons and ergoregion of the black hole. 
In Section \ref{sec:3}, we examine how the black hole spin parameter $a$, electric charge $Q$, magnetic charge $v$, NUT charge $N_g$, and gauge coupling $g$ affect the event horizons, ergoregion, and circular geodesics of the black hole in question.
In Section \ref{sec:4}, we present a detailed analysis of energy extraction via the CA mechanism, exploring the relevant parameters and phase spaces, and computing both the energy extraction rate and reconnection efficiency. Our main results are summarized in Section \ref{sec:5}. Throughout this work, we use geometrized units with $c = G = 1$.

\section{Rotating Dyonic Black Holes in \texorpdfstring{$\mathcal{N}=2, U(1)^2$}{N=2, U(1)\texttwosuperior} Gauged Supergravity}\label{sec:2}

In this section, we analyze the geodesics of a rotating dyonic black hole solution within a four-dimensional $\mathcal{N} = 2, U(1)^2$ gauged supergravity.
This metric was originally derived in \cite{Chow2014} and further analyzed in \cite{Flathmann2016}.
This solution generalizes the Kerr-Newman-AdS metric by including a NUT parameter and allowing for dyonic charges.
We present a summary of this derivation below.

The derivation begins with the gauged Lagrangian of the generic form,
\begin{align}
	\mathcal{L}_{\text{gauged}} = \mathcal{L}_4 + g^2 V(\Phi_A) \star 1,
\end{align} 
\noindent where $g$ is the gauge coupling constant, $V(\Phi_A)$ is the scalar potential depending on the scalar fields $\Phi_A$, and $\mathcal{L}_4$ is the ungauged Lagrangian, which is obtained after a Kaluza-Klein reduction of the 11-dimensional supergravity theory on $\mathcal S^7$ to give a gauged $\mathcal N{=}8$ theory with $SO(8)$ symmetry \cite{Cvetic1999}. 

Two truncations are then made; the first is a consistent truncation that brings the theory to a $\mathcal{N}{=}2$, $U(1)^4$ theory, and the second brings the theory to $U(1)^2$ by setting $A^1{=}A^4$, $\tilde{A}^2{=}\tilde{A}^3$, and turning off all but one active dilaton $\varphi$ and one active axion. 

The resulting action is
	\begin{equation}
		\begin{split}
			S = \frac{1}{\kappa} \int d^4x \sqrt{-g} \Bigg[
			& R - \frac{1}{2} \partial_\mu \varphi \partial^\mu \varphi
			- \frac{1}{2} e^{2\varphi} \partial_\mu \chi \partial^\mu \chi
			- \frac{1}{2} e^{-\varphi} F^1_{\mu\nu} F^{1\mu\nu}
			+ \frac{1}{4} \chi \epsilon^{\mu\nu\rho\sigma} F^1_{\mu\nu} F^1_{\rho\sigma}
			\\[2pt]
			& - \frac{1}{1 + \chi^2 e^{2\varphi}}
			\left(
			\frac{1}{2} e^\varphi F^2_{\mu\nu} F^{2\mu\nu}
			+ \frac{1}{4} \chi e^{2\varphi}
			\epsilon^{\mu\nu\rho\sigma} F^2_{\mu\nu} F^2_{\rho\sigma}
			\right)
			+ g^2(4 + e^\varphi + e^{-\varphi} + \chi^2 e^\varphi)
			\Bigg],
		\end{split}
		\label{eq:N2U2action}
	\end{equation}
where $\kappa = 16\pi G$.

The resulting rotating black hole solution found in \cite{Chow2014} is,
\begin{align}
	ds^2 &= -\frac{\hat{R}_g}{\hat{W}} \left( d\hat{t} - \hat{a} \frac{\hat{u}_1\hat{u}_2}{\hat{a}^2} d\hat{\phi} \right)^2 + \frac{\hat{W}}{\hat{R}_g} d\hat{r}^2 \nonumber \\
	&+ \frac{\hat{U}_g}{\hat{W}} \left( d\hat{t} - \hat{a} \frac{\hat{r}_1\hat{r}_2}{\hat{a}^2} d\hat{\phi} \right)^2 + \frac{\hat{W}}{\hat{U}_g} d\hat{u}^2 ,\label{eq:local_metric}
\end{align}
where the metric functions are
\begin{align}
	\hat{W} &= \hat{r}_1\hat{r}_2 + \hat{u}_1\hat{u}_2 \label{eq:Gamma_hat}, \\
	\hat{U}_g(\hat{u}) &= -\hat{u}^2 + 2\hat{n}\hat{u} + \hat{a}^2 + g^2\hat{u}_1\hat{u}_2(\hat{u}_1\hat{u}_2 - \hat{a}^2) \label{eq:U_g}, \\
	\hat{R}_g(\hat{r}) &= \hat{r}^2 - 2\hat{m}\hat{r} + \hat{a}^2 + g^2\hat{r}_1\hat{r}_2(\hat{r}_1\hat{r}_2 + \hat{a}^2).
\end{align}

Here, $\hat{R}_g(\hat r)$ and $\hat{U}_g(\hat u)$ are quadratic polynomials in $\hat r$ and $\hat u$ modified by the gauge coupling $g^2$, while $\hat r_{1,2}$ and $\hat u_{1,2}$ are linear functions of $\hat r$ and $\hat u$ shifted by the parameters $\Delta_{\hat r}$ and $\Delta_{\hat u}$. 
In this form, the coordinates are somewhat inconvenient: the metric rotates at infinity rather than approaching standard AdS coordinates, and $\hat\phi$ is not canonically normalized. 
Because we want to define the energy with respect to a static observer at infinity, we switch to a regularized, asymptotically AdS coordinate system.

The black hole solution can also be expressed in a coordinate system that is more relevant to this study. The transformation to the new coordinates regularizes the hatted coordinates and asymptotically moves them towards AdS \cite{Chow2014}.
To facilitate the coordinate transformation, the shift parameters $\Delta$ are rewritten in terms of their sums ($\Sigma_\Delta$) and differences ($\Delta_\Delta$) as
\begin{subequations}
\begin{align}
	\Sigma_{\Delta_r} &= \frac{1}{2}(\Delta_{\hat{r}_1} + \Delta_{\hat{r}_2}), \quad \Delta_{\Delta_r} = \frac{1}{2}(\Delta_{\hat{r}_2} - \Delta_{\hat{r}_1}), \\
	\Sigma_{\Delta_u} &= \frac{1}{2}(\Delta_{\hat{u}_1} + \Delta_{\hat{u}_2}), \quad \Delta_{\Delta_u} = \frac{1}{2}(\Delta_{\hat{u}_2} - \Delta_{\hat{u}_1}).
\end{align}
\end{subequations}
Using these definitions, and substituting the individual shifted coordinates into their respective products, yields a difference of squares
\begin{subequations}
\begin{align}
	\hat{r}_1\hat{r}_2 &= (\hat{r} + \Sigma_{\Delta_r})^2 - (\Delta_{\Delta_r})^2 \label{eq:r_product}, \\
	\hat{u}_1\hat{u}_2 &= (\hat{u} + \Sigma_{\Delta_u})^2 - (\Delta_{\Delta_u})^2 \label{eq:u_product}.
\end{align}
\end{subequations}
The Griffiths-Podolsk\'y transformation is then used to obtain an asymptotically AdS coordinate system by applying a transformation to map $(\hat{t}, \hat{r}, \hat{u}, \hat{\phi})$ to the new coordinates $(t, r, p, \bar{\phi})$ via
\begin{subequations}
\begin{align}
	\hat{r} &= \beta r - \Sigma_{\Delta_r} \implies d\hat{r} = \beta dr \label{eq:r_trans}, \\
	\hat{u} &= \beta(N_g + ap) - \Sigma_{\Delta_u} \implies d\hat{u} = a\beta dp, \\
	\hat{\phi} &= \frac{a}{\hat{a}\beta^3} \bar{\phi} \implies d\hat{\phi} = \frac{a}{\hat{a}\beta^3} d\bar{\phi}, \\
	\hat{t} &= \frac{t}{\beta} + \frac{\hat{a}^2 + \Delta_{\Delta_u}^2 - (N_g + a)^2\beta^2}{a\beta^3} \bar{\phi} \nonumber \\
	&\implies d\hat{t} = \frac{dt}{\beta} + \frac{\hat{a}^2 + \Delta_{\Delta_u}^2 - (N_g + a)^2\beta^2}{a\beta^3} d\bar{\phi}.
\end{align}
\end{subequations}

By applying the radial and angular transformations to the products in Eqs. (\ref{eq:r_product}) and (\ref{eq:u_product}), and substituting these back into the definition of $\hat{W}$ (Eq. \ref{eq:Gamma_hat}), the transformed conformal factor is found:
\begin{equation}
	\hat{W} = \beta^2 \left[ r^2 + (N_g + ap)^2 - v^2 \right] \equiv \beta^2 W, \label{eq:Gamma_transformed}
\end{equation}
where $v^2 = \beta^{-2}(\Delta_{\Delta_r}^2 + \Delta_{\Delta_u}^2)$ is the charge parameter, and $W = r^2 + (N_g + ap)^2 - v^2$.

We substitute the transformed conformal factor $\hat{W} = \beta^2 W$ and $d\hat{r}^2 = \beta^2 dr^2$ directly into the radial term $\frac{\hat{W}}{\hat{R}_g} d\hat{r}^2$ to obtain $\frac{\beta^4 W}{{R}_g} dr^2$, where  $\hat{R}_g = \beta^4 R_g$ is defined so that the radial term matches the standard form.

Next, we transform the angular component $\hat{U}_g$ into a regularized polynomial $P_g$ by substituting Eq. (\ref{eq:Gamma_transformed}) and the new angular differential $d\hat{u} = \beta a dp$ directly into the original angular metric term $\frac{\hat{W}}{\hat{U}_g} d\hat{u}^2$, which yields $\frac{\beta^4 a^2 W}{\hat{U}_g} dp^2$. To match the standard generalized Plebański–Demiański metric form of $\frac{W}{P_g} dp^2$, we must define the relationship $\hat{U}_g = \beta^4 a^2 P_g$.
To arrive at the generalized metric form, we substitute the transformed time and angular differentials, $d\hat{t}$ and $d\hat{\phi}$, into the squared differential components $(d\hat{t} - \dots d\hat{\phi})^2$ of the original metric (Eq. \ref{eq:local_metric}). By expanding these squared terms and carefully factoring out the new $dt$ and $d\bar{\phi}$ variables, we systematically grouped the remaining constants into the geometric coefficients as follows:
\begin{align}
	ds^2 &= -\frac{R_g}{W} \Big( dt - [2N_g(1 - p) + a(1 - p^2)]d\bar{\phi} \Big)^2 \nonumber \\
	&+ \frac{P_g}{W} \Big( a dt - [r^2 - v^2 + (N_g + a)^2]d\bar{\phi} \Big)^2 \\
	&+ W \left( \frac{dr^2}{R_g} + \frac{dp^2}{P_g} \right) ,\label{eq:generalized_metric}
\end{align}
where $R_g$ is a quartic polynomial in $r$, and the behavior of the angular coordinate $p$ is governed by the quartic polynomial $P_g(p) = a_0 + a_1p + a_2p^2 + a_3p^3 + a_4p^4$. The coefficients are derived by substituting the transformed coordinate $\hat{u} = \beta(N_g + ap) - \Sigma\Delta_u$ into the original angular polynomial $\hat{U}_g(\hat{u})$ (Eq. \ref{eq:U_g}), expanding it, and collecting terms by powers of $p$
\begin{subequations}
\begin{align}
	a_0 &= a^{-2}(k - N_g^2 \epsilon + 2nN_g + g^2N_g^4), \\
	a_1 &= 2a^{-1}(n - N_g\epsilon + 2g^2N_g^3), \\
	a_2 &= 6g^2N_g^2 - \epsilon, \\
	a_3 &= 4ag^2N_g, \\
	a_4 &= g^2a^2.
\end{align}
\end{subequations}

To ensure that the solution corresponds to a regular, spherically symmetric black hole, we require the angular polynomial $P_g(p)$ to vanish at the poles $p=\pm1$. The conditions $P_g(\pm1)=0$ imply

\begin{align}
 a_0 + a_2 + a_4 = 0, \qquad a_1 + a_3 = 0.   
\end{align}

Fixing the overall normalization by choosing $a_0=1$, these relations determine the remaining parameters $k$, $\epsilon$, and $n$ as

\begin{align}
k &= (a^{2} - N_{g}^{2})(1 + 3 g^{2} N_{g}^{2}), \label{eq:k-def}\\
\epsilon &= 1 + g^{2}\bigl(a^{2} + 6 N_{g}^{2}\bigr), \label{eq:eps-def}\\
n &= N_{g}\bigl[1 - g^{2}(a^{2} - 4 N_{g}^{2})\bigr]. \label{eq:n-def}
\end{align}

Finally, to obtain the physical metric, we substitute the angular parameter $p = \cos\theta$. This determines the angular polynomial as $P_g = \sin^2\theta(1 - 4ag^2N_g\cos\theta - a^2g^2\cos^2\theta) = \sin^2\theta \, \Theta_g$. We then rescale the azimuthal coordinate using the parameter $\Xi$, mapping $\bar{\phi} = \phi / \Xi$, where $\Xi = 1 - a^2g^2 - 4aN_gg^2$.

Let $B = r^2 - v^2 + (N_g + a)^2$ and $A = 4N_g\sin^2(\theta/2) + a\sin^2\theta$. Utilizing the definitions of $A$ and $B$, it becomes clear that $B - aA = W$. Substituting everything back into Eq. (\ref{eq:generalized_metric}) yields the final regularized metric result
\begin{align}\label{eqn:ds2}
	ds^2 &= -\frac{\Delta_g}{B - aA} \left( dt - \frac{A}{\Xi} d\phi \right)^2 + \frac{B - aA}{\Delta_g} dr^2 \nonumber \\
	&+ \frac{\Theta_g a^2 \sin^2\theta}{B - aA} \left( dt - \frac{B}{a\Xi} d\phi \right)^2 + \frac{B - aA}{\Theta_g} d\theta^2.
\end{align}

The auxiliary functions $A$ and $B$ and the scaling factor $\Xi$ incorporate the rotation parameter $a$, NUT charge $N_g$, and gauge coupling $g$
\begin{align}
	A &= a \sin^2\theta + 4N_g \sin^2(\theta/2), \\
	B &= r^2 + (N_g + a)^2 - v^2, \\
	\Xi &= 1 - 4a N_g g^2 - a^2 g^2. 
\end{align}
The metric functions $\Delta_g$ and $\Theta_g$, which determine the horizon structure and angular dependence, respectively, are defined as
\begin{align}\label{eqn:Deltag}
	\Delta_g &= r^2 + a^2 - 2Mr + Q^2 - N_g^2 + g^2 \left[ r^4 + (a^2 + 6N_g^2 - 2v^2)r^2 + 3N_g^2 (a^2 - N_g^2) \right],\\
\label{eqn:Thetag}
	\Theta_g &= 1 - a^2 g^2 \cos^2\theta - 4a^2 N_g \cos\theta.
\end{align}

\section{Geodesics of a rotating dyonic black hole in the $N=2$, $U(1)^2$ gauged supergravity}\label{sec:3}

Using regularized spacetime, we next studied its geometric and dynamical properties. We show how the parameters ($a, Q, g, N_g, v$) deform the event horizon and ergoregion, and subsequently derive the geodesic equations to locate the innermost stable circular orbits (ISCOs) and photon spheres for the test particles.

The specific covariant metric components $g_{\mu\nu}$ can be identified from the line element above as
\begin{align}
\label{eqn:gtt}
g_{tt} &= -\frac{\Delta_g - a^2 \Theta_g \sin^2\theta}{B - aA},\\[0.5em]
\label{eqn:phiphi}
g_{\phi\phi} &= \frac{\sin^2\theta}{\Xi^2 (B - aA)} 
\left[ \Theta_g (r^2 + (N_g+a)^2 - v^2)^2 - \Delta_g A^2 \right],\\[0.5em]
\label{eqn:tphi}
g_{t\phi} &= \frac{1}{\Xi (B - aA)} 
\left[ \Delta_g A - \Theta_g a \sin^2\theta (r^2 + (N_g+a)^2 - v^2) \right]. 
\end{align}
This solution is characterized by six distinct physical parameters: the black hole mass $M$, spin parameter $a$, electric charge $Q$, magnetic charge $v$, NUT charge $N_g$, and gauge coupling constant $g$. Parameter $g$ is related to the inverse length scale of the Anti-de Sitter (AdS) space ($g = 1/L$). The function $\Delta_g$ recovers the standard Kerr-Newman horizon structure \cite{Newman1965} when $v=0$, $g=0$, and $N_g=0$, but introduces quartic corrections $g^2 r^4$ in the gauged supergravity context.

Because the viability of the energy extraction mechanisms depends critically on the volume and topology of the ergosphere, we next analyze the horizon and ergoregion structures for the black hole spacetime under consideration. The event horizons are identified by the roots of the radial function given by
\begin{align}\label{eqn:Delta_g}
	\Delta_g &= r^2 + a^2 - 2Mr + Q^2 - N_g^2 + g^2 \left[ r^4 + (a^2 + 6N_g^2 - 2v^2)r^2 + 3N_g^2 (a^2 - N_g^2) \right] = 0.
\end{align}

When $g=N_g=v=0$, this expression reduces to the standard Kerr-Newman horizon structure, $r^2 + a^2 - 2Mr + Q^2=0$.
However, the defining feature of this supergravity solution is the quartic correction term scaled by the square of the gauge coupling ($g^2r^4$). At large radii, this term dominates and yields an asymptotically AdS geometry \cite{Abbott1982}. Physically, the $g^2 r^4$ term acts as an AdS-like confining potential, corresponding to a positive thermodynamic pressure in gauged supergravity \cite{Kastor2009,HawkingPage1983,Dolan2011,Kubiznak2017}. As $g$ increases, this effective pressure reshapes the global geometry and shifts the outer horizon to smaller radii, tightening the region where stable orbits and energy extraction are possible.

\begin{figure*}%[!htbp]
	\centering
	\includegraphics[width=0.99\linewidth]{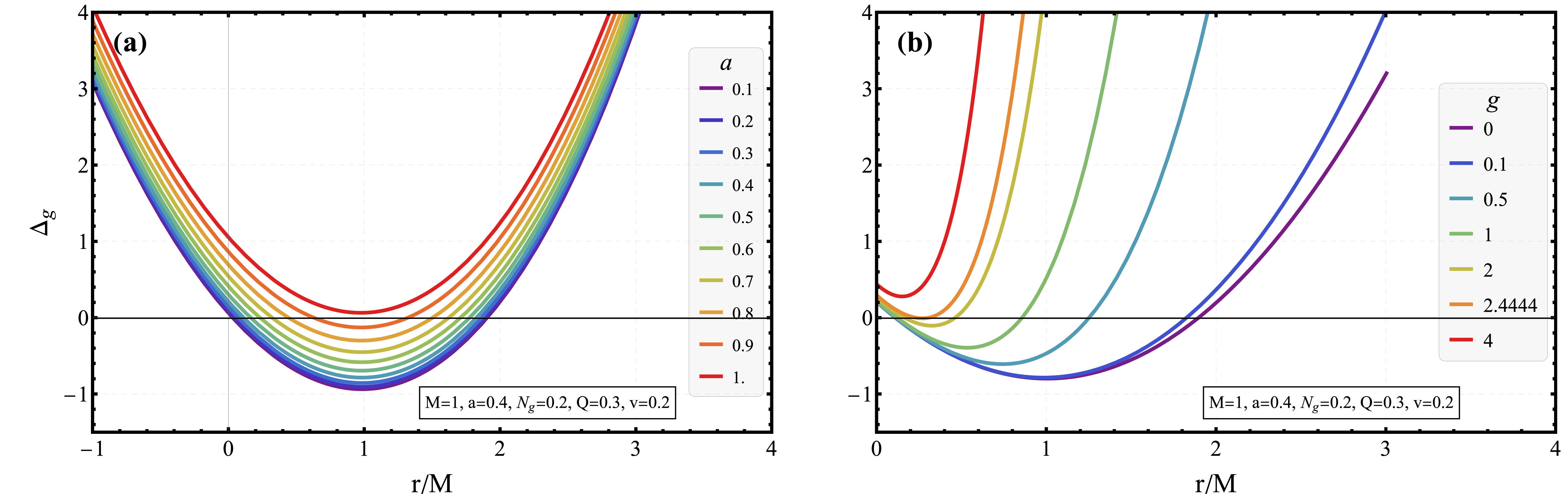}
    \caption{The metric function $\Delta_g(r)$ is shown for fixed $M = 1, N_g = 0.2, Q = 0.3$, and $v = 0.2$. Left: Varying spin $a$ at fixed $g = 0.1$. Right: Varying coupling $g$ at fixed $a = 0.4$. In both panels, the zeros of $\Delta_g$ (where $\Delta_g = 0$) mark the horizon. As the varied parameter increases, the spacetime passes from a two-horizon black hole, through an extremal case with merged horizons, to a naked singularity with no horizons.}
	\label{fig:Combined_Delta_Panel}
\end{figure*}

Figure~\ref {fig:Combined_Delta_Panel}(a) shows the changes in the horizon structure as we vary the spin parameter $a$ from $0.1$ to $1.0$, keeping everything else fixed at $M = 1, g = 0.2, N_g = 0.2, Q = 0.3$, and $v = 0.2$. At low spins, $\Delta_g$ crosses zero twice for $r > 0$, indicating that the black hole has both an inner and an outer horizon, which is the usual picture for a non-extremal black hole. Increasing $a$ pushes the two horizons toward each other; the inner one moves outward, the outer one contracts, and at some critical spin, they coincide, giving the extremal case. Beyond this point, $\Delta_g$ remains positive, and the horizons disappear entirely, leaving behind a naked singularity.
The nonlinear electrodynamic coupling, $g$, behaves similarly as shown in Figure~\ref{fig:Combined_Delta_Panel}(b), which we vary over ${0, 0.1, 0.5, 1, 2, 2.4444, 4}$ while holding $M = 1, a = 0.4, N_g = 0.2, Q = 0.3$, and $v = 0.2$. Setting $g = 0$ recovers the ordinary Kerr–Newman solution with two well-separated horizons; that is, as $g$ increases, the $g^2 r^4$ term lifts the $\Delta_g$ profile upward, squeezing the horizons together. At $g \approx 2.4444$, the curve is tangent to the zero axis, marking the extremal configuration, and at $g=4$ it is strictly positive, and no horizons remain. Figures ~\ref {fig:Combined_Delta_Panel}(a) and ~\ref {fig:Combined_Delta_Panel}(b) tell essentially the same story; varying either $a$ or $g$ drives the black hole through the transition from two horizons to extremality to a naked singularity. 

To study the geodesics of this black hole, we consider a test particle with the Lagrangian 
\begin{align}
	\mathcal{L}=\frac{1}{2}g_{\mu\nu}\dot{x}^{\mu}\dot{x}^{\nu},
\end{align}
where the derivative is with respect to the proper time $\tau$. Detailed derivations and analytical solutions of these geodesic equations for this specific supergravity background, including a comprehensive classification of bound and escape orbits, were established by \cite{Flathmann2016}. Timelike geodesics satisfy the normalization condition $g_{\mu\nu}u^\mu u^\nu = -1$.
Because spacetime is stationary and axisymmetric, the Killing symmetries associated with $t$ and $\phi$ imply two constants of motion. These are the energies per unit mass:
\begin{equation}
	E \equiv -p_\mu \xi_{(t)}^\mu = -p_t,
\end{equation}
and azimuthal angular momentum per unit mass:
\begin{equation}
	L \equiv p_\mu \xi_{(\phi)}^\mu = p_\phi,
\end{equation}
where $p_\mu = g_{\mu\nu}u^\nu$ is the four-momentum of the particle. In terms of the metric components, these conserved quantities are given by
\begin{equation}
	p_t = g_{tt}\dot t + g_{t\phi}\dot\phi = -E,
	\qquad
	p_\phi = g_{t\phi}\dot t + g_{\phi\phi}\dot\phi = L.
\end{equation}
In these rotating geometries the Hamilton–Jacobi equation is separable because the metric admits a non‑trivial Killing tensor and a corresponding Carter constant, as first shown by Carter \cite{Carter1968}. The same hidden symmetry exists in our supergravity solution; therefore, particle motion remains completely separable.

Solving this linear system for $\dot t$ and $\dot\phi$ in terms of $(E,L)$ and the metric functions yields
\begin{equation}
	\dot t =
	\frac{g_{\phi\phi}\,E + g_{t\phi}\,L}
	{g_{t\phi}^2 - g_{tt}g_{\phi\phi}},
	\qquad
	\dot\phi =
	-\frac{g_{t\phi}\,E + g_{tt}\,L}
	{g_{t\phi}^2 - g_{tt}g_{\phi\phi}}.
\end{equation}
In this supergravity background, the geodesic equations are integrable because the metric admits a nontrivial conformal Killing–Yano 2‑form, from which one can construct the usual quadratic constant of motion. This structure was worked out explicitly in \cite{Chow2014}, and it guarantees the separability of the Hamilton–Jacobi equation.

We restrict the analysis to motion within the equatorial plane, defined by
$\theta = \frac{\pi}{2} \ \text{and}\ \dot\theta = 0$. 
A dimensionless parameter $b$ is introduced via the relation $L = b\,E$.
It is then convenient to define an effective potential $V_{\mathrm{eff}}(r)$ by $V_{\mathrm{eff}}(r) = -\frac{1}{2}\,\dot r^2$, so that the radial motion can be written in the familiar form $\frac{1}{2}\,\dot r^2 + V_{\mathrm{eff}}(r) = 0$. Circular orbits at radius $r = r_0$ are characterized by $\dot r = 0 \ \text{and}\ \ddot r = 0$.
Therefore, the effective potential conditions are
\begin{equation}
	V_{\mathrm{eff}}(r_0) = 0,
	\qquad
	\frac{dV_{\mathrm{eff}}}{dr}(r_0) = 0.
\end{equation}
The first condition is automatically satisfied at the level of the radial equation, so the nontrivial circularity condition is
$\frac{dV_{\mathrm{eff}}}{dr} = 0 $
The stability of a circular orbit at $r = r_0$ is controlled by the second derivative of the effective potential $\frac{d^2V_{\mathrm{eff}}}{dr^2}(r_0)$.
A stable circular orbit corresponds to a local minimum of $V_{\mathrm{eff}}$, that is, $d^2V_{\mathrm{eff}}/dr^2 > 0$, while an unstable circular orbit corresponds to a local maximum, that is, $d^2V_{\mathrm{eff}}/dr^2 < 0$. The ISCO is located at the boundary between stable and unstable circular orbits, where the curvature vanishes
\begin{equation}
	\frac{d^2V_{\mathrm{eff}}}{dr^2}\bigg|_{r=r_{\mathrm{ISCO}}} = 0.
\end{equation}

We define the ISCO radius $r_{\mathrm{ISCO}}$ as the root of $V_{\mathrm{eff}}^{\rm ISCO}(r;M,a,g,N_g,Q,P) = 0$
corresponds to a physical, timelike circular orbit outside the event horizon.

We used Newton’s method to find this root. With an initial guess $r_0$, we iterate
\begin{align}
	r_{n+1} = r_n - \frac{V_{\mathrm{eff}}^{\rm ISCO}(r_n)}{\partial_r V_{\mathrm{eff}}^{\rm ISCO}(r_n)},
\end{align} 
until convergence.

The left panel of Figure~\ref{fig:iscophoton} shows how the radius of the innermost stable circular orbit, $r_{\rm ISCO}/M$, depends on the black hole spin $a/M$ for several values of the gauge coupling $g$ in the rotating dyonic $\mathcal{N}=2, U(1)^2$ black hole. Each colored curve corresponds to a different coupling, $g = 0.1, 0.3, 0.5, 0.7,$ and $0.9$. For any fixed $g$, $r_{\rm ISCO}$ decreases steadily as the spin increases, so higher spin pulls the ISCO closer to the horizon, as in Kerr spacetime. At a given spin, we find numerically that a larger $g$ yields a smaller ISCO radius; over the range shown, the red curve ($g=0.9$) lies innermost, while the blue curve ($g=0.1$) lies outermost. The termination of the curves at finite $a/M$ indicates the approach to the extremal bound, beyond which our solutions no longer admit a regular ISCO for this coupling. Overall, increasing either the spin or gauge coupling moves the ISCO inward and compresses the region of stable circular motion, which can make accretion and energy extraction more efficient in this model. This trend is somewhat reminiscent of the ISCO tightening observed for charged particles in Kerr–Newman spacetime \cite{Hackmann2013}, although here it is induced purely by the scalar fields and the AdS confining potential of the supergravity background.
\begin{figure*}%[!htbp]
	\centering
	\begin{minipage}[b]{0.48\linewidth}
		\centering
		\includegraphics[height=5.3cm]{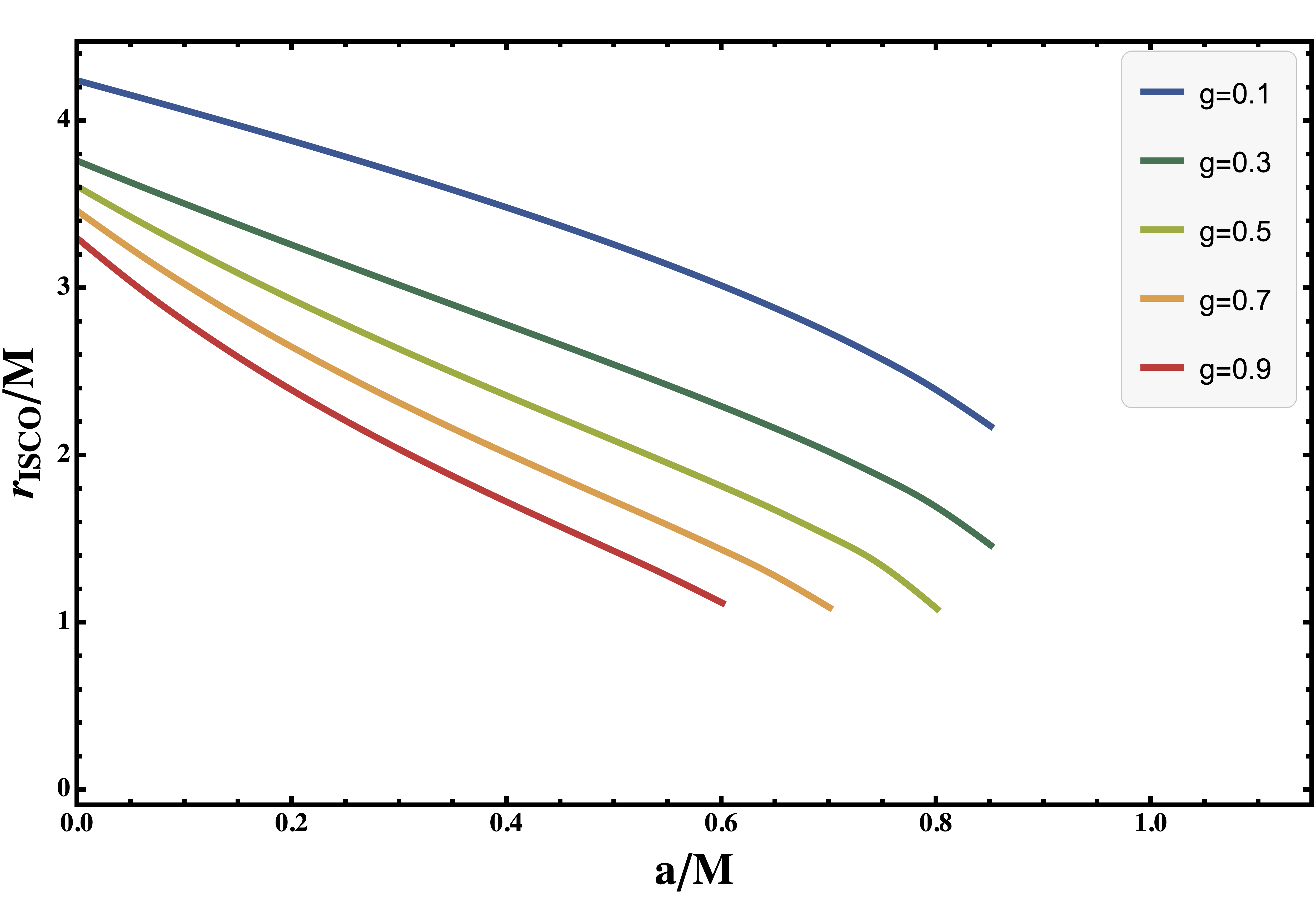}
	\end{minipage}
	\hfill
	\begin{minipage}[b]{0.48\linewidth}
		\centering
		\includegraphics[height=5.3cm]{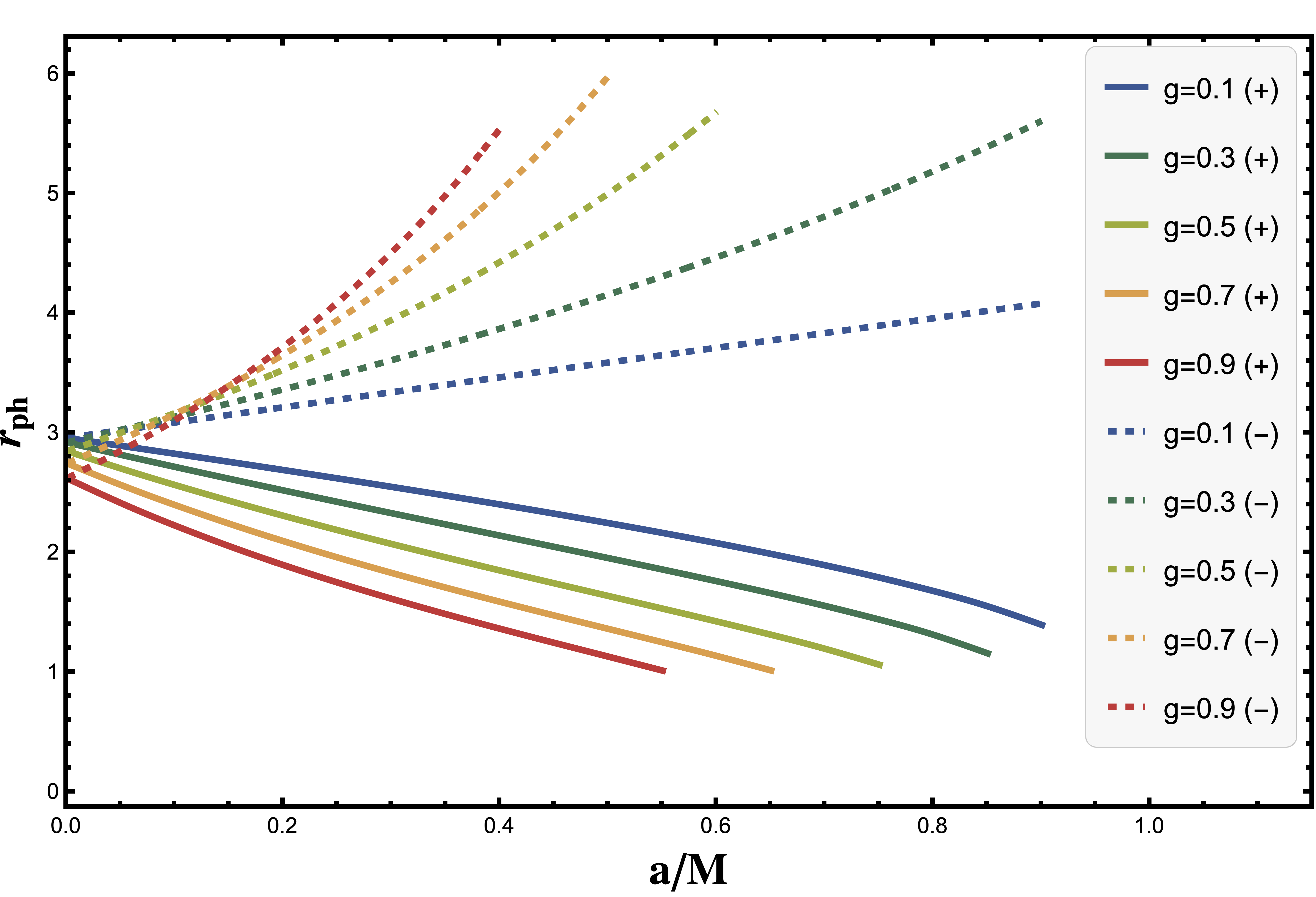}
	\end{minipage}
	\caption{The equatorial ISCO (left) and photon sphere (right) radii of bound photon orbits in the modified SUGRA regime. The split in the photon sphere plot illustrates the frame-dragging effect on prograde ($+$) vs. retrograde ($-$) null geodesics. We adopt the parameter values $M=1, N_g=0.2, Q=0.3, v=0.2$. The different curves in both panels correspond to gauge couplings $g=0.1,0.3,0.5,0.7,$ and $0.9$.}
	\label{fig:iscophoton}
\end{figure*}

For null geodesics (photon trajectories), the line element is zero: \(ds^2 = 0\). If we now confine the motion to the equatorial plane (\(\theta = \pi/2\) with \(\dot{\theta} = 0\)) and consider only circular orbits (\(\dot{r} = 0\)), the metric (Eq.~\ref{eqn:ds2}) takes the simpler form
\begin{align}g_{tt} \dot{t}^2 + 2 g_{t\phi} \dot{t} \dot{\phi} + g_{\phi\phi} \dot{\phi}^2 = 0.\end{align}
Expressing $\dot{t}$ and $\dot{\phi}$ in terms of the conserved energy $E$ and angular momentum $L$ leads to a quadratic equation for the impact parameter $b \equiv L/E$. Solving this equation, we obtain $b(r)$

\begin{align}b(r) = \frac{-g_{t\phi} \pm \sqrt{g_{t\phi}^2 - g_{tt} g_{\phi\phi}}}{g_{tt}}.\end{align}
The circularity condition further requires $\ddot{r}=0$, which is satisfied at the extrema of the impact parameter, $\partial_r b(r) = 0$. This condition identifies the radius of the photon sphere, $r_{ph}$, which corresponds to the unstable equilibrium point at which the radial effective potential reaches its maximum value. 

The right panel of Figure~\ref{fig:iscophoton} shows the radius of the circular photon orbits $r_{ph\pm}$ as a function of $a/M$. For each $g$, the prograde branch (solid lines) moves inward, while the retrograde branch (dashed lines) moves outward as the spin increases, producing a clear split between the co‑rotating and counter‑rotating null orbits. This spin–induced separation of photon spheres, familiar from Kerr, translates into a corresponding asymmetry of the black hole shadow (see Refs. ~\cite{Bambi2009, Sharma2021} for general discussions related to spacetimes.
\begin{figure}%[!htbp]
	\centering
    \includegraphics[width=0.5\linewidth]{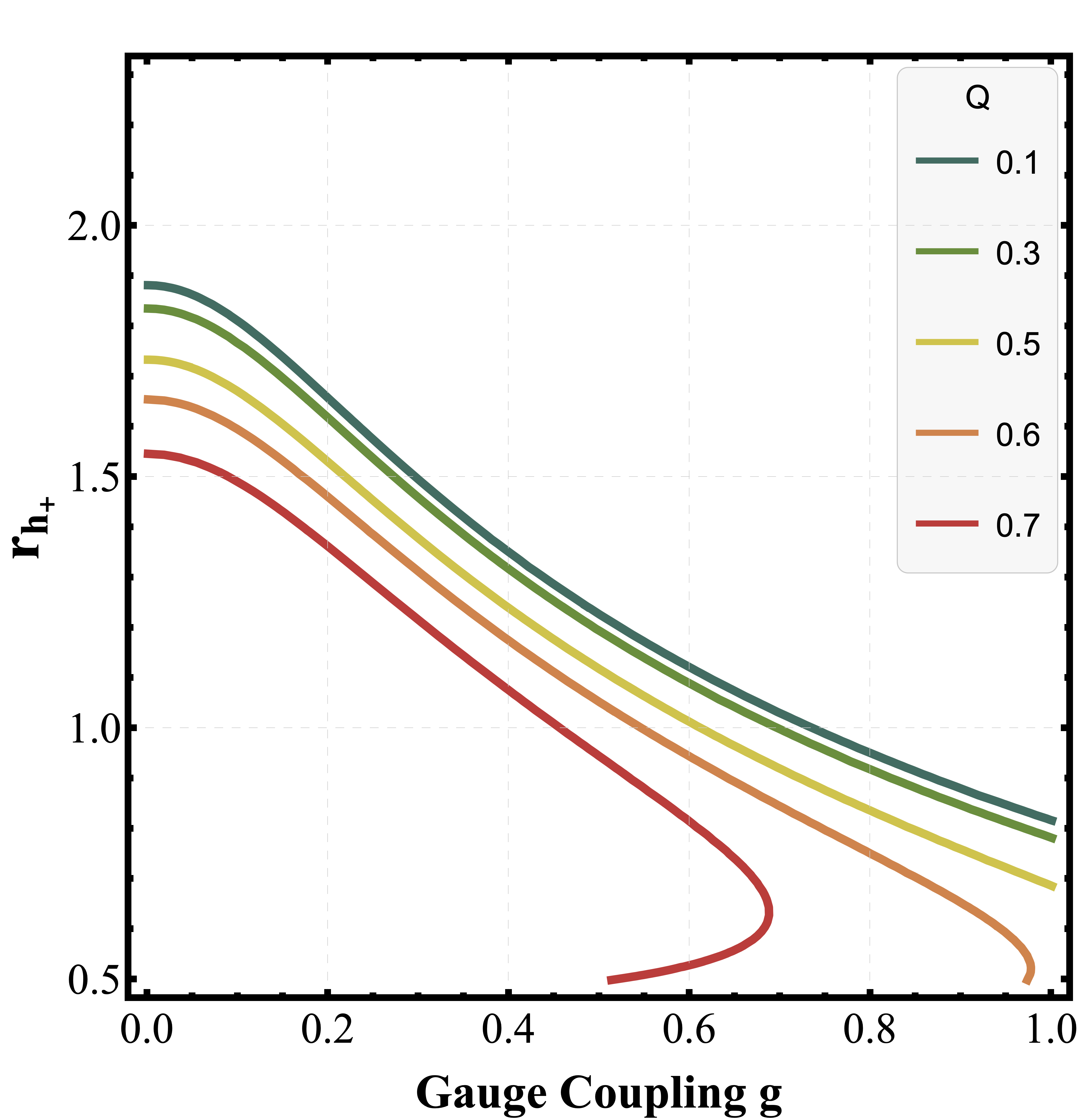}
 \caption{Event horizon radius $r_{h_+}$ versus gauge coupling $g$ for rotating dyonic black holes with $M = 1, a = 0.5, v = 0.2, N_g = 0.2$ and electric charges $Q = 0.1\text{–}0.7$ (green to red). Because of the $g^2 r^4$ term, $r_{h_+}$ decreases monotonically with $g$, more steeply at larger $Q$ due to stronger scalar–charge coupling.}
    \label{fig:horizonplusVSgvaryQ}
\end{figure}
Figure~\ref{fig:horizonplusVSgvaryQ} shows the dependence of the outer horizon radius $r_{h+}$ on the gauge coupling $g$ for a rotating dyonic $\mathcal{N}=2,\,U(1)^2$ gauged supergravity black hole, with different curves corresponding to $Q = 0.1, 0.3, 0.5, 0.6,$ and $0.7$. As $g$ increases, the horizon radius decreases for all charges, and for $Q=0.1,0.3,0.5,0.6$, this decrease is monotonic. At fixed $g$, a larger $Q$ yields a smaller $r_{h+}$, with the most highly charged case ($Q=0.7$) yielding the smallest horizon. For the highest charge shown ($Q=0.7$), the curve $r_{h+}(g)$ develops a turning point around $g \simeq 0.6$ and then terminates, signaling an extremal bound in the $(Q,g)$ plane beyond which no regular outer horizon exists; the milder‑charge curves extend to larger $g$ before terminating.

For the ergoregion structure, we analyze the static limit defined by 
\begin{align}\label{eqn:gtt0}
	g_{tt} =\Delta_g - a^2 \Theta_g \sin^2\theta = 0. 
\end{align}
We define the outer static limit radius $r_{SL}$ as the largest positive root of Eq.~\ref{eqn:gtt0}. The physical thickness of the ergosphere in the equatorial plane is given by $\delta r = r_{SL} - r_{H+}$.
Unlike the horizon equation (Eq. \ref{eqn:Delta_g}), this surface depends explicitly on the angular coordinate $\theta$ via the function $\Theta_g$ (defined in Eq. \ref{eqn:Thetag}), leading to the non-spherical topology of the ergosphere. 

Figure~\ref{fig:ergosphere_structure} shows the meridional ($x–z$) cross‑sections of the event horizon (solid green) and the static limit (dashed orange) for the rotating dyonic $\mathcal{N}=2,\ U(1)^2$ gauged supergravity black hole. The spin was fixed at $a=0.8$. The top-left panel shows the Kerr limit $(Q=g=N_g=v=0)$, where both the horizon and the static limit are nearly spherical, and the ergoregion is relatively thick. In the other three panels, we fix $Q=0.3,\ N_g=0.2,\ v=0.2$ and increase the gauge coupling from $g=0.2$ to $0.4$ and $0.5$. Both the horizon and the static limit move inward as $g$ increases, and the radial gap between them visibly shrinks; therefore, the ergoregion becomes progressively thinner.
This behavior reflects the $g$-dependence of the horizon and static limit equations. The horizon is set by $\Delta_g(r)=0$, whose large $r$ behavior is controlled by the quartic AdS term $g^2 r^4$ in Eq.~\eqref{eqn:Delta_g}. The static limit satisfies $g_{tt}=0$ and is also influenced by the scalar‑charge terms in the $g^2(a^2+6N_g^2-2v^2)r^2$ contribution to $\Delta_g$ and $\Theta_g$ \cite{Chow2014}. Together, these effects drive both surfaces inward as $g$ increases, compressing the ergoregion. The same parameters that control this compression also limit the allowed black hole configurations, since the real roots of $\Delta_g(r)=0$ and the condition $\Xi = 1 - 4 a N_g g^2 - a^2 g^2 > 0$ restrict how large $a$, $g$, and the charges can be before the horizon or the Lorentzian signature are lost \cite{Hawking1999,Kostelecky1996}.
\begin{figure*}%[!htbp]
	\centering
	\includegraphics[width=0.9\linewidth]{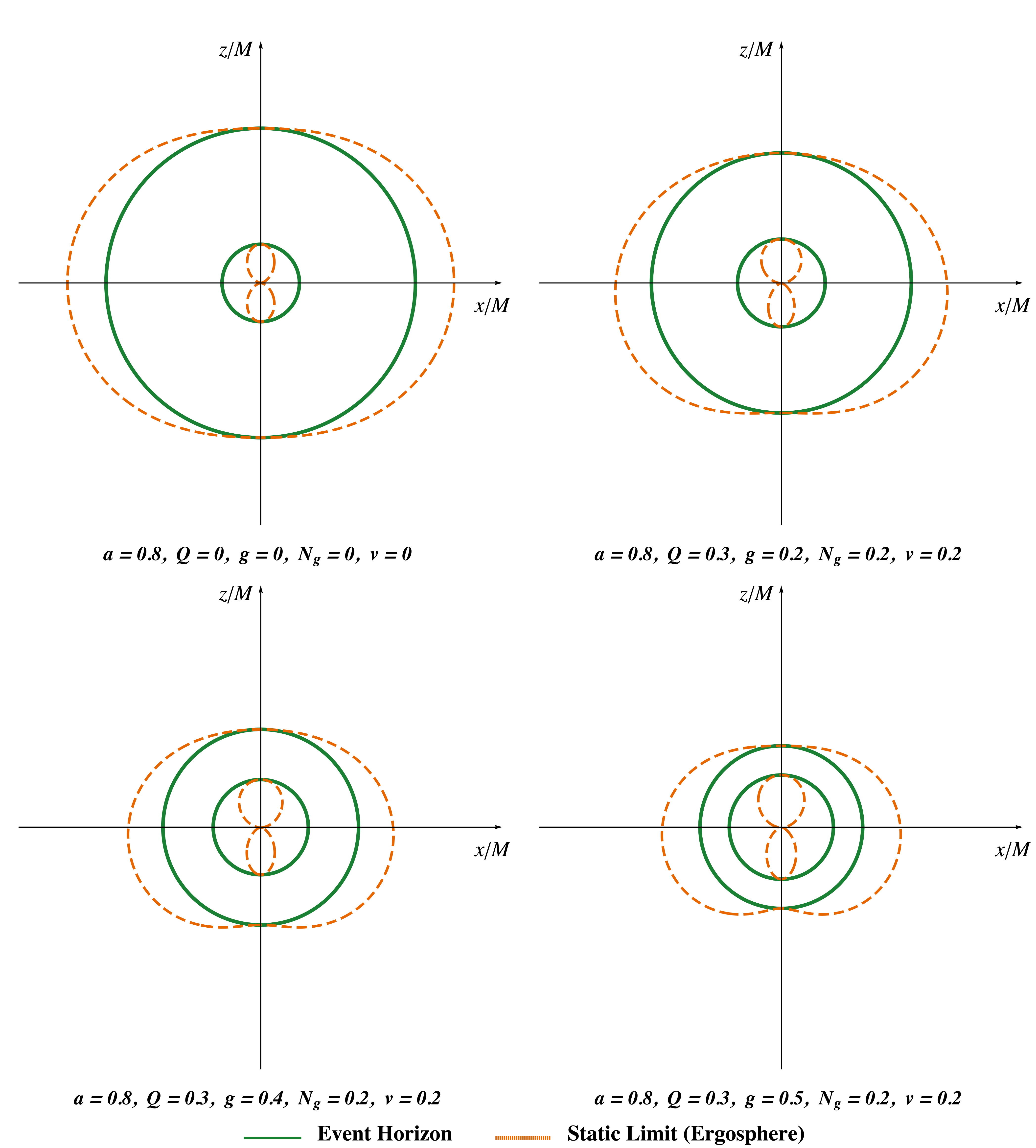}
    \caption{Cross-sections of horizons (green) and static limits (orange dashed) for rotating dyonic black holes with $M=1, a=0.8$ in $\mathcal{N}=2,\ U(1)^2$ gauged supergravity. The top left panel is for Kerr ($g=0$). Middle and right plots are for $Q=0.3, v=0.2, N_g=0.2$ at $g=0.2, 0.4, 0.5$. The ergosphere widens and peaks at $g=0.4$, then contracts at $g=0.5$, showing a non-monotonic response to the scalar potential.}
	\label{fig:ergosphere_structure}
\end{figure*}

From the condition $\Xi > 0$, we derive an exact analytical bound for the gauge coupling in terms of the spin and NUT charge
\begin{align}
	g < \frac{1}{\sqrt{a(a + 4N_g)}}.
\end{align} 

For the parameters used in Figure~\ref{fig:ergosphere_structure} ($a=0.8, N_g=0.2$), this yields a critical value of $g_{crit} \approx 0.88$, which is consistent with the numerical breakdown observed in this study.
Consequently, high-spin black holes ($a=0.8$) encounter stability limits at lower gauge couplings ($g_{crit} \approx 0.9$) than slowly rotating black holes. In contrast, increasing the electric charge $Q$ drives the system closer to extremality. As $g$ increases, the effective width of the ergosphere is shifted, and the geometry ceases to be well-defined for couplings above $g_{crit} \approx 1.2$, which is essentially independent of the charge. In this regime, the efficiency of energy extraction is controlled by the competition between supersymmetry breaking, governed by $g$, and rotational effects.

With these geometric bounds in hand, in Section~\ref{sec:3} we examine whether the available ergosphere volume can sustain the plasma magnetization needed to realize negative-energy states and, hence, efficient energy extraction.

\begin{figure*}%[!htbp]
	\centering
	\includegraphics[width=0.99\linewidth]{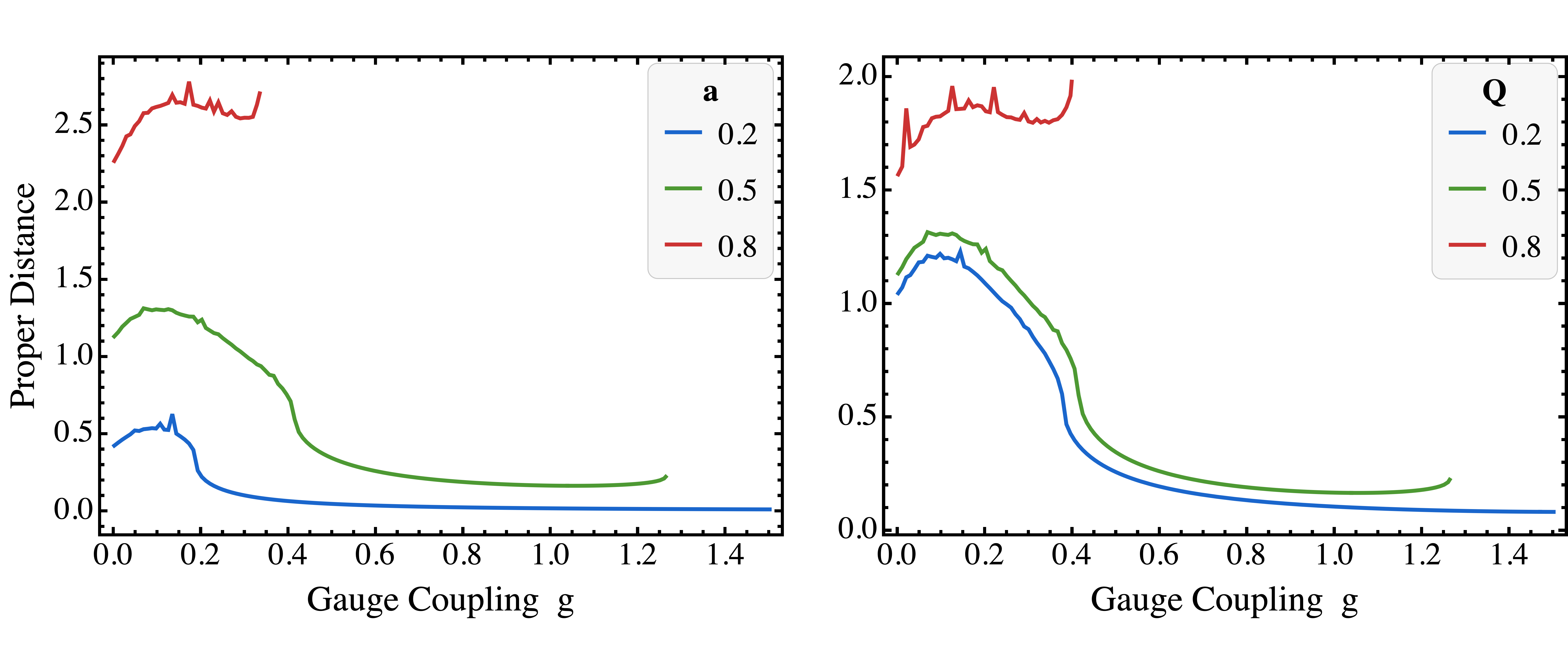}
    \caption{Proper equatorial ergoregion thickness $\ell(r_H \to r_{\rm SL})$versus gauge coupling $g$ for $M=1, N_g=0.2, v=0.2$. Left: varying spin $a=0.2,0.5,0.8$ at fixed $Q=0.5$. Right: Varying charge $Q=0.2,0.5,0.8$ at fixed $a=0.5$. In both panels, a larger $a$ or $Q$ thickens the ergoregion at small $g$ but narrows the range of $g$ over which a regular horizon–ergoregion configuration exists.}
	\label{fig:properD-combined}
\end{figure*}
The proper radial distance between $r_H$ and $r_{SL}$ is  
\begin{align}
      \ell(g,a) \;=\; \int_{r_H}^{r_{SL}} \sqrt{\big|g_{rr}(r)\big|}\,dr. 
   \end{align}
Using this definition, we evaluate the equatorial proper distance between the horizon and the static limit as a function of the gauge coupling and other black hole parameters. Figure~\ref{fig:properD-combined} displays how the proper ergoregion thickness, $\ell(r_H \to r_{\rm SL})$, varies with $g$ for fixed $M=1, Q=0.5, N_g=0.2$ and $v=0.2$. The three curves correspond to spins $a=0.2, 0.5, 0.8$. For the slowly rotating case ($a=0.2$), the ergoregion is relatively thin, and its thickness rapidly decreases to nearly zero as $g$ increases, indicating that the static limit approaches the horizon. For $a=0.5$ and $a=0.8$, the ergoregion is substantially thicker at small $g$. In the $a=0.5$ case, the proper thickness decreases as $g$ increases, while for $a=0.8$ a sizeable ergoregion exists only for small $g$; beyond a critical value of $g$ the solution no longer admits a regular horizon–static–limit pair, and the curve terminates.
The right panel shows the effect of changing the electric charge at a fixed spin $a=0.5$. The same qualitative pattern appears: for all three charges, the proper thickness $\ell(r_H \to r_{\rm SL})$ is largest at small $g$ and decreases as $g$ increases. A larger $Q$ produces a thicker ergoregion at any given $g$ and restricts the allowed range of gauge coupling: the high-charge curve ($Q=0.8$) terminates at a smaller $g$ than the low-charge one, reflecting that increasing $Q$ drives the spacetime toward extremality and eventually destroys the regular horizon–ergoregion structure. In practice, increasing either $a$ or $Q$ narrows the region in the parameter space where the ergoregion has a moderate size and confines efficient energy extraction to a smaller domain in $(g,a,Q)$.

\section{Energy Extraction from a rotating dyonic black hole in the $N=2$, $U(1)^2$ gauged supergravity}\label{sec:4}
\subsection{Magnetic Reconnection via the Comisso-Asenjo Process}

In this section, we investigate the energy extraction via the Comisso-Asenjo process for the rotating dyonic black hole in the $N=2$, $U(1)^2$ gauged supergravity and analyze how our parameter space influences the efficiency of the energy extraction process, as in \cite{Comisso2021}. 
To simplify the calculations, it is common to perform calculations in the local zero-angular-momentum-observer (ZAMO) frame \cite{Bardeen1972}, a locally non-rotating frame in which spacetime appears flat (Minkowskian) for the observer. 
The tetrads relating the global frame to the ZAMO frame are 
\begin{align}
	d\hat{t} &= \alpha dt, \\
	d\hat{x}^i &= \sqrt{g_{ii}} dx^i - \alpha \beta^i dt,
\end{align}
where $i$ represents the spatial indices $r, \theta, \phi$. The lapse function, $\alpha$, is given by
\begin{align}
	\alpha = \sqrt{-g_{tt} + \frac{g_{t\phi}^2}{g_{\phi\phi}}}
	= \sqrt{\frac{\Delta_g \Theta_g (B - a A)}{B^2 \Theta_g - A^2 \Delta_g}}.
\end{align}
The shift vector, $\beta^i$, is defined by using the frame drag angular velocity $\omega^\phi$
\begin{equation}
	\beta^\phi = \frac{\sqrt{g_{\phi\phi}} \omega^\phi}{\alpha} 
	= \sqrt{\frac{\Delta_g \Theta_g (B - a A)}{B^2 \Theta_g - A^2 \Delta_g}},
\end{equation}
\begin{align}
	\omega^{\phi}=-g_{t\phi}/g_{\phi\phi} = \frac{\Xi  (A \Delta_g -a B \Theta_g)}{A^2 \Delta_g - B^2 \Theta_g}.
\end{align}
In this analysis, only the azimuthal component $\beta^\phi$ is relevant ($\beta^r = \beta^\theta = 0$), which accounts for the frame-dragging effect and ensures that the velocity is measured relative to that of the non-rotating observer. The contravariant vectors in the ZAMO frames can be transformed to the Boyer-Lindquist coordinates by $\hat{b}^0=\alpha b^0$ and $\hat{b}^i=\sqrt{g_{ii}}b^i-\alpha \beta^ib^0$, while covectors transform by $\hat{b}_0=b_0/\alpha+\Sigma^3_{i=1}(\beta^i/\sqrt{g_{ii}})b_i$ and $\hat{b}_i=b_i/\sqrt{g_{ii}}$.

For the plasma fluid element to maintain a circular orbit in the equatorial plane ($\theta = \pi/2$), we require
$$ \partial_r g_{\mu\nu} u^\mu u^\nu = 0. $$
Note that the only nonzero components of the four-velocity $u$ are the time ($u^t = \dot{t}$) and azimuth ($u^\phi\dot{\phi}$) components. Given that angular velocity is defined by $\Omega = \frac{d\phi}{dt} = \frac{u^\phi}{u^t}$, we obtain the Keplerian angular velocity in Boyer-Lindquist coordinates from
\begin{align}
	\Omega_K = \frac{-\partial_r g_{t\phi} \pm \sqrt{(\partial_r g_{t\phi})^2 - (\partial_r g_{tt})(\partial_r g_{\phi\phi})}}{\partial_r g_{\phi\phi}}.
\end{align}

Using the above transformations, the Keplerian velocity of the co-rotating bulk plasma in the ZAMO frame can be found by 
\begin{align}
	\hat{v}_K=\frac{d\hat{x}^{\phi}}{d\hat{t}}=\frac{\sqrt{g_{\phi\phi}}}{\alpha}{\Omega_K}-\beta^{\phi}.
\end{align}
Assuming a one-fluid approximation for the bulk plasma surrounding the black hole (where the magnetic reconnection actually occurs), we can describe the magnetic reconnection process using the contribution to the total energy-momentum tensor ($T^{\mu\nu}_b$), which is given as \cite{Comisso2021}
\begin{align}\label{eqn:T}
	T^{\mu\nu}_b = p g^{\mu\nu} + \omega U^{\mu} U^{\nu} + F^{\mu}{}_{\delta} F^{\nu\delta} - \frac{1}{4} g^{\mu\nu} F^{\rho\delta} F_{\rho\delta}.
\end{align}
Here $T_{b}^{\mu\nu}$ is distinct from the energy-momentum tensor $T^{s}_{\mu\nu}$ sourcing the black hole; since we treat the spacetime as fixed and the bulk plasma as dynamically negligible, the electromagnetic field tensor $F_{\mu\nu}$ in the above equation receives contributions from the bulk plasma alone, leaving the black hole charge $Q$ unaffected.
The first part of Eq. \ref{eqn:T}, $pg^{\mu\nu} + \omega U^\mu U^\nu$, describes the plasma as a perfect fluid, where $p$ is the proper pressure and $U^\mu$ is the four-velocity. The term $\omega$ represents the enthalpy density, defined as $\omega = \rho + u + p$, where $\rho$ is the rest mass density and $u$ is the internal energy density.
Here, we assume that the spacetime is fixed and that the bulk plasma does not affect the black hole charge $Q$. Thus, the electromagnetic field tensor $F^{\mu\nu}$ in the second part of Eq. (~\ref{eqn:T}) represents only the magnetic field lines inside the plasma (those twisting and breaking during reconnection), which are distinct from the static electric field generated by the black hole's charge $Q$.

Next, we evaluate the ``energy-at-infinity" density by a stationary observer at infinity using a time-like Killing vector $\xi_\nu$ \cite{Comisso2021}:
\begin{align}\label{eqn:energy}
	e^{\infty}\equiv n_{\mu}J^{\mu}=n_{\mu}T^{\mu\nu}_b\xi_{\nu}=-\alpha g_{\mu 0}T^{\mu 0}_b=\alpha \hat{e}+\alpha \beta^\phi \hat{P}^\phi,  
\end{align}
where $\xi_\nu = (\partial_t)_\nu = (\partial_t, 0, 0, 0)$, $n_\nu$ is the unit vector normal to the time-like hyperfaces, $\hat{e}$ is the total local energy density, and $\hat{P}^\phi$ is the local azimuthal momentum. The total energy density $\hat{e}$ that combines the energy from the matter (plasma) and the electromagnetic fields, and the azimuthal momentum density $\hat{P}_\phi$ as measured by a local observer in the ZAMO frame, are given as
\begin{align}
	\hat{e} &= w\hat{\gamma}^2 - p + \frac{\hat{B}^2 + \hat{E}^2}{2},\nonumber \\
	\hat{P}^\phi &= w\hat{\gamma}^2\hat{v}^\phi + (\boldsymbol{\hat{B}} \times \boldsymbol{\hat{E}})^\phi.
\end{align}
where $w\hat{\gamma}^2\hat{v}^\phi$ represents the mechanical momentum of the moving plasma in the $\phi$ direction, and $\hat{v}^\phi$ is the azimuthal velocity of the plasma. The outflow Lorentz factor $\hat{\gamma}$ for the ZAMO frame is defined using the three-velocity $\hat{v}^i$ as 
$$ \hat{\gamma} = \hat{U}^0 = \left[ 1 - \sum_{i=1}^3 (d\hat{v}^i)^2 \right]^{-1/2}. $$
Adopting the same approach as in \cite{Comisso2021}, we define the magnetic field components using the electromagnetic field tensor $\hat{F}$ and the Levi-Civita tensor $\epsilon^{ijk}$ as 
$ \hat{B}^i = \frac{\epsilon^{ijk}\hat{F}_{jk}}{2} $
and the electric field $\hat{E}^i$ using the metric $\eta^{ij}$ and the field tensor as
$ \hat{E}^i = \eta^{ij}\hat{F}_{j0} = \hat{F}_{i0} $. The hydrodynamic component of the "energy-at-infinity" density $e^\infty_{hyd}$ in the ZAMO frame is given by the following equations:
\begin{subequations}
\begin{align}
	\hat{e}_{hyd} &= T^{\hat{0}\hat{0}}_{hyd} = \omega (U^{\hat{0}})^2 + p \eta^{\hat{0}\hat{0}} = \omega \hat{\gamma}^2 - p \label{eq:e-hydro} \\
	\hat{P}^{\phi}_{hyd} &= T^{\hat{0}\hat{\phi}}_{hyd} = \omega U^{\hat{0}} U^{\hat{\phi}} = \omega \hat{\gamma}^2 \hat{v}_\phi.  \label{eq:P-hydro}
\end{align}
\end{subequations}

The total "energy-at-infinity" density $e^\infty$ defined in Eq. \ref{eqn:energy} can be separated into 
hydrodynamic and electromagnetic  
$e^\infty = e^\infty_{hyd} + e^\infty_{em}$, where
\begin{align}
	e^\infty_{hyd} &= \alpha (\omega \hat{\gamma}^2 - p) + \alpha \beta^\phi (\omega \hat{\gamma}^2 \hat{v}^\phi), \\
	e^\infty_{em} &= \alpha \hat{e}_{em} + \alpha \beta^\phi \hat{P}^{\phi}_{em}.
\end{align}
We follow the same assumption made in \cite{Comisso2021} that the magnetic reconnection process acts as an efficient engine that converts the available magnetic energy into kinetic energy. Consequently, the electromagnetic term in the energy density at infinity, $e^\infty$, effectively vanishes. Also, assuming that the expelled plasma is incompressible and adiabatic, the system's energy balance 
is determined entirely by the hydrodynamic component, leading to the following expression: 
\begin{align}\label{eqn:einfy}
	e^\infty_{hyd} =   \alpha\hat{\gamma}w (1 + \beta^\phi\hat{v}^\phi)  - \frac{\alpha p}{\hat{\gamma}},
\end{align}
as in \cite{Koide2008}.
This expression for the energy balance depends on the local ZAMO Lorentz factor $\hat{\gamma}$ and azimuthal velocity $\hat{v}^\phi$. 
Extreme frame-dragging by a rapidly spinning black hole leads to complex, time-dependent, nonlinear dynamics, including magnetic reconnection. These dynamics stretch and reorient magnetic field lines in ways that depend on the large-scale field configuration and the spin of the black hole.
It is necessary to analyze the reconnection process from a local rest frame $x^{\mu'}$ that co-rotates with the bulk plasma at Keplerian angular velocity. Within this local frame, the arbitrary trajectory of the reconnecting magnetic field lines is parameterized by the orientation angle $\xi$. As defined below, this angle represents the ratio of the radial to azimuthal outflow velocities of the escaping plasma
\begin{align}
	\xi = \arctan(\frac{v'^1_{out}}{v'^3_{out}}).
\end{align}

The macroscopic energy equations were evaluated using Eq. \ref{eqn:einfy}; the local kinematics of the reconnection outflow must also be transformed into the ZAMO frame. This translation was achieved by applying the relativistic-velocity addition formula. The system is modeled using two reference frames, the ZAMO frame and a local rest frame $x^{\mu'}$ that co-rotates with the bulk plasma at the Keplerian velocity $\hat{v}_K$. Within this local frame, the azimuthal component of the expelled plasma is parameterized by its local outflow velocity ($v_{out}$) and arbitrary orientation angle of the reconnecting magnetic field lines ($\xi$), resulting in a local azimuthal speed of $\pm v_{out} \, \cos\xi$ for the corotating ($+$) and counterrotating ($-$) plasma outflows
	\begin{align}\label{eqn:energyhydroinfty}
		e^\infty_{hyd,\pm} &= \alpha\hat{\gamma}_K \left[ (1+\hat{v}_K\beta^\phi)\gamma_{out}w \pm cos\xi (\hat{v}_K+\beta^\phi)\gamma_{out}v_{out}w - \frac{p}{(1 \pm \cos\xi\hat{v}_K v_{out})\gamma_{out}\hat{\gamma}_K^2} \right].
	\end{align}
Assuming a thin current sheet, the final outflow speed is entirely dictated by the asymptotic upstream plasma magnetization ($\sigma_0 = B_0^2/w_0$). In a highly magnetized regime ($\sigma_0 \gg 1$), the magnetic energy efficiently determines the ejection speed, forcing the outflow Lorentz factor $\gamma_{out}$ to scale directly with magnetization, yielding 
\begin{align}
	v_{out}&\approx\left (\frac{\sigma_0}{1+\sigma_0}\right )^{1/2}, \label{eqn:vout}\\
	\gamma_{out}&=(1-v_{out}^2)^{-1/2}\approx(1+\sigma_0)^{1/2}. \label{eqn:gammaout}
\end{align}
Adding the Keplerian velocity of the local frame to the plasma's local azimuthal velocity relativistically yields the total ZAMO azimuthal velocity as follows \cite{Comisso2021}
\begin{align}
	\hat{v}^{\phi}_{\pm} = \frac{\hat{v}_K \pm v_{out} \,cos\xi}{1 \pm \hat{v}_K v_{out} \, cos\xi}.
\end{align}

To derive the final asymptotic energy, the magnetization-dependent kinematics from Eq. ~\ref{eqn:vout} and Eq. ~\ref{eqn:gammaout} must be substituted back into the intermediate energy expression in Eq. ~\ref{eqn:energyhydroinfty}. To resolve the proper pressure term $p$, the fluid was modeled as a relativistically hot plasma governed by a polytropic index of $\Gamma = 4/3$, which establishes a simple pressure-to-enthalpy ratio of $p/\omega = (\Gamma-1)/\Gamma = 1/4$. By dividing the entire expression by the proper enthalpy density $\omega$, the equation is normalized to yield the specific energy at infinity per unit enthalpy $\epsilon^\infty_\pm$, resulting in the final expression for the energy density at infinity per enthalpy, which leads to as in \cite{Comisso2021}
\begin{align}\label{eqn:energyfinal}
	\begin{split}
		\epsilon^\infty_\pm = \alpha\hat{\gamma}_K \Big[ &(1+\beta^\phi\hat{v}_K)(1+\sigma_0)^{1/2} \pm \cos\xi (\beta^\phi+\hat{v}_K)\sigma_0^{1/2} \\ 
		&\quad - \frac{(1+\sigma_0)^{1/2} \mp \cos\xi\hat{v}_K\sigma_0^{1/2}}{4\hat{\gamma}_K^2(1+\sigma_0 - \cos^2\xi\hat{v}_K^2\sigma_0)} \Big].
	\end{split}
\end{align}

\subsection{Constraints on Cutoff Plasma Magnetization $\sigma_0$ \& Parameter Space}
Figure~\ref{fig:energyvssigma} shows how the energies at infinity of the two Comisso–Asenjo branches, $\epsilon_\pm^{\infty}$, depend on the plasma magnetization, $\sigma_0$. The horizontal axis represents $\sigma_0$, and the vertical axis represents $\epsilon_\pm^{\infty}$. Solid curves give the “minus” branch $\epsilon_-^{\infty}$, dashed curves the “plus” branch $\epsilon_+^{\infty}$, and different colors correspond to gauge couplings $g = 0.2, 0.3, 0.5, 0.7$. For all values of $g$, the dashed curves $\epsilon_+^{\infty}$ increase monotonically with $\sigma_0$ and remain positive, with larger $g$ producing a higher $\epsilon_+^{\infty}$ at a given magnetization. The solid curves $\epsilon_-^{\infty}$ show the opposite trend for small and moderate $g$: for $g=0.2,0.3$ (and more weakly for $g=0.5$) they decrease with $\sigma_0$ and become more negative as the plasma becomes more magnetized, while for larger $g$ (e.g., $g=0.7$) they stay close to zero and remain positive. Thus, strong negative energy states $\epsilon_-^{\infty}<0$, required for efficient Comisso–Asenjo energy extraction, are favored by high magnetization at small–intermediate $g$; as $g$ increases, the negative branch is pushed upward, and the parameter range with $\epsilon_-^{\infty}<0$ shrinks.
\begin{figure}%[!htbp]
	\centering
	\includegraphics[width=0.7\linewidth]{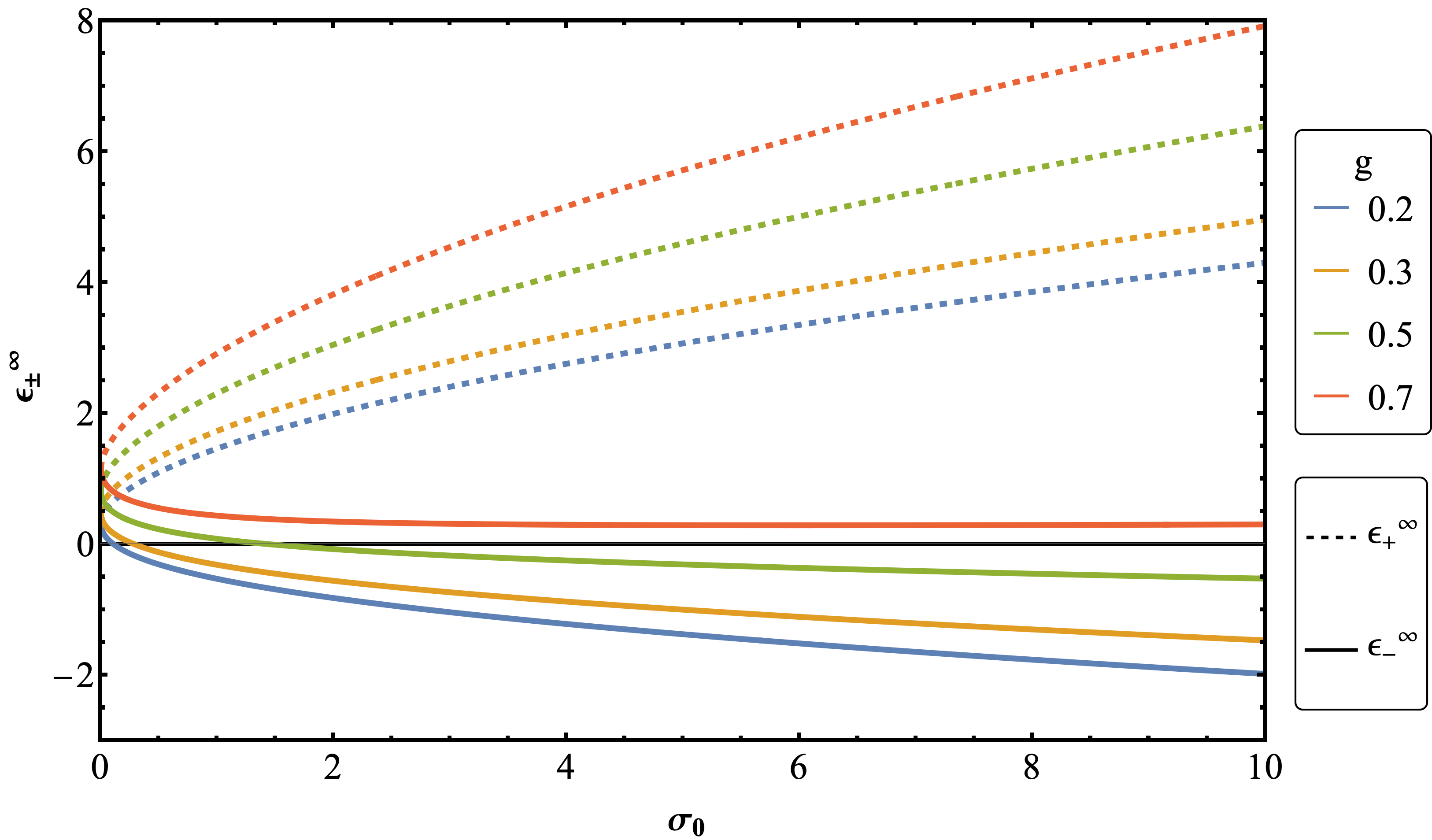}
	\caption{ Energy-at-infinity density per enthalpy, $\epsilon^{\infty}_{+}$ (solid lines) and $\epsilon^{\infty}_{-}$ (dashed lines) as a function of plasma magnetization $\sigma_0$ for various coupling $g$ values.}
	\label{fig:energyvssigma}
\end{figure}

\begin{figure*}%[!htbp]
	\centering
	\includegraphics[width=0.48\linewidth]{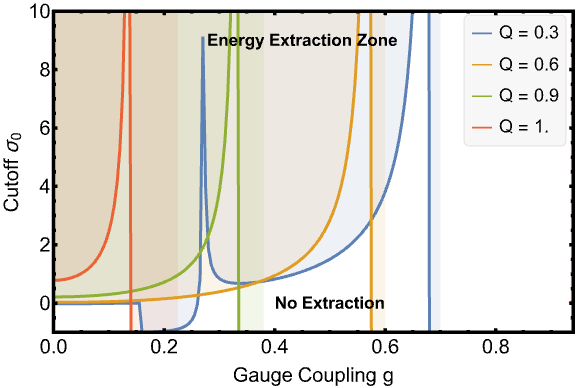}
	\hfill
	\includegraphics[width=0.48\linewidth]{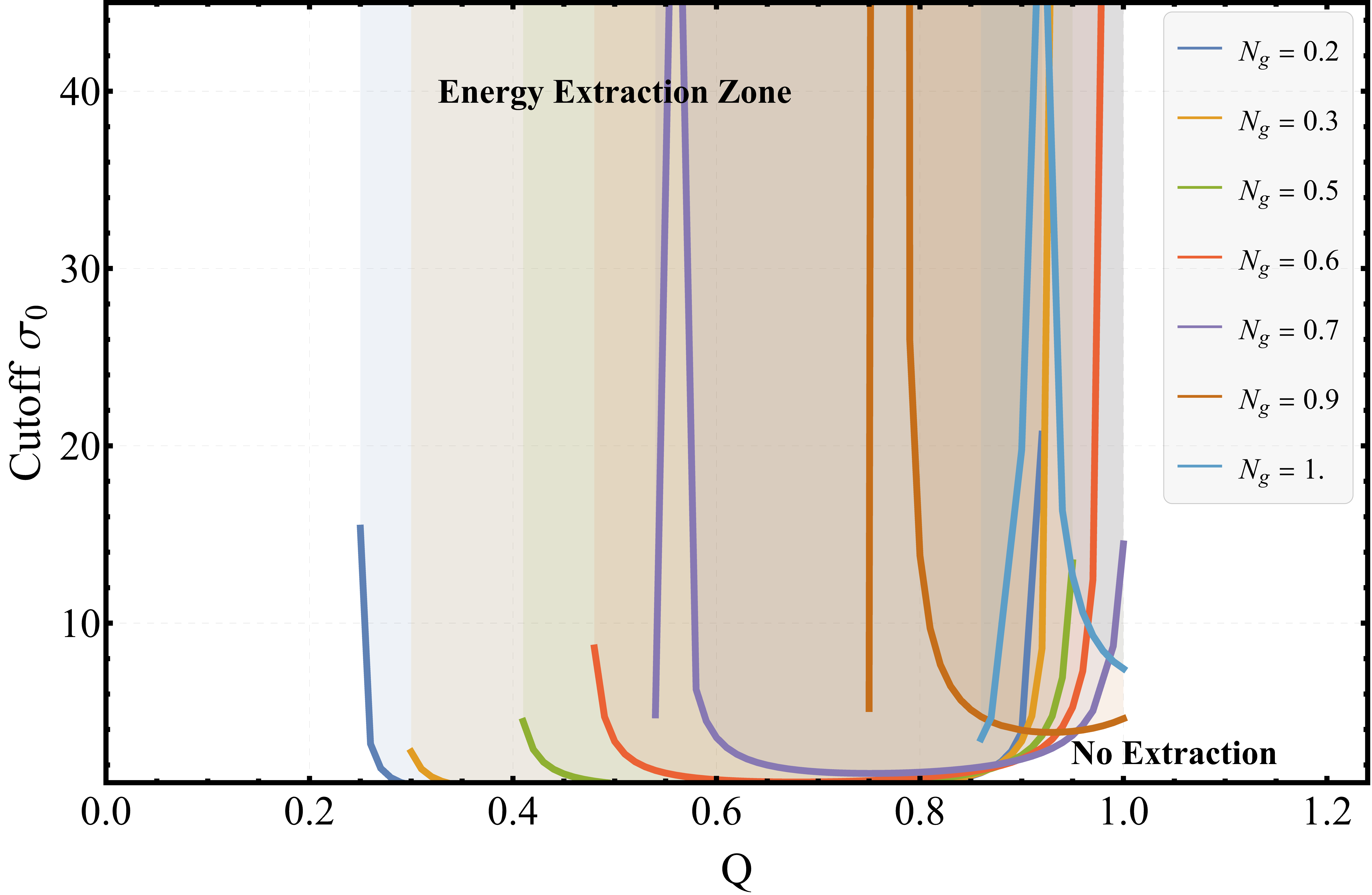}
	\caption{Cutoff magnetization $\sigma_0$ required for energy extraction via magnetic reconnection.  
		Left: $\sigma_0$ as a function of the gauge coupling \(g\) for selected charges $Q$, with $M = 1$, $a = 0.9$, $N_g = 0.2$, $v = 0.2$, and $\xi = 0$.  
		Right: $\sigma_0$ as a function of the black hole charge $Q$ for several values of the NUT parameter $N_g$ for the same background parameters. In both panels, the shaded region indicates the parameter domain in which the energy of the escaping plasmoid satisfies $e^{\infty}_{-} < 0$, corresponding to successful energy extraction, while unshaded regions mark parameter values for which reconnection cannot extract energy.}
	\label{fig:sigmavsgQ}
\end{figure*}
The left panel of Figure~\ref{fig:sigmavsgQ} shows the cutoff magnetization $\sigma_0$ as a function of the gauge coupling $g$ for fixed charges $Q=0.3,0.6,0.9,1.0$. Each curve gives, for that $Q$, the value of $\sigma_0$ at which the negative energy branch $\epsilon_-^{\infty}$ crosses zero. For $\sigma_0$ below the curve, $\epsilon_-^{\infty}>0$ and CA extraction is not possible (“No Extraction”), whereas for $\sigma_0$ above the curve, $\epsilon_-^{\infty}<0$ and negative energy inflow is allowed (“Energy Extraction Zone”). The near-vertical spikes mark the critical couplings $g_{\rm crit}(Q)$ where the horizon becomes extremal or the negative energy branch ceases to exist.

The right panel displays the analogous cutoff curves $\sigma_0(Q)$ for fixed $N_g$ at $g=0.3$. For each $N_g$, extraction operates only when $\sigma_0 > \sigma_0^{\rm cutoff}(Q,N_g)$; the unshaded region below the curves corresponds to $\epsilon_-^{\infty}>0$, whereas the shaded region above them admits extraction. Tables~\ref{tab:tab1} and \ref{tab:tab2} list the representative numerical values of $\sigma_0^{\rm cutoff}(g,Q)$ and $\sigma_0^{\rm cutoff}(Q,N_g)$ that define these curves. The 3D phase diagram in Figure ~\ref{fig:Sigma_PhaseDiagram_3D} summarizes the full dependence of the cutoff $\sigma_0$ on \(g\) and \(Q\); the surface \(\epsilon_-^{\infty}=0\) represents the CA activation threshold, with parameter points above (below) the surface allowing (forbidding) negative energy inflow.  

From Table~\ref{tab:tab1}, we see that for small–to–intermediate $g$ and moderate charge, the required magnetization is modest, $\sigma_0=\mathcal{O}(0.1\!-\!1)$, with a shallow “valley” around $Q\simeq0.5\text{--}0.6$ where the threshold is lowest. At larger $g$ or for near‑extremal charges, the cutoff grows rapidly (e.g., $\sigma_0\sim80$ for $g=0.65, Q=0.4$), indicating that only very strongly magnetized plasma could extract energy in that regime. Table~\ref{tab:tab2} shows a similar behavior as $Q$ and $N_g$ increase. 

The main takeaway is that efficient CA extraction exists only in a bounded region of $(g, Q,\sigma_0)$. Increasing the gauge coupling or pushing the charge toward extremality raises the required magnetization and shrinks the viable parameter space; thus, highly charged, strongly gauged black holes admit energy extraction only in a much smaller and more finely tuned region.

\begin{figure*}%[!htbp]
	\centering
	\includegraphics[width=0.7\textwidth]{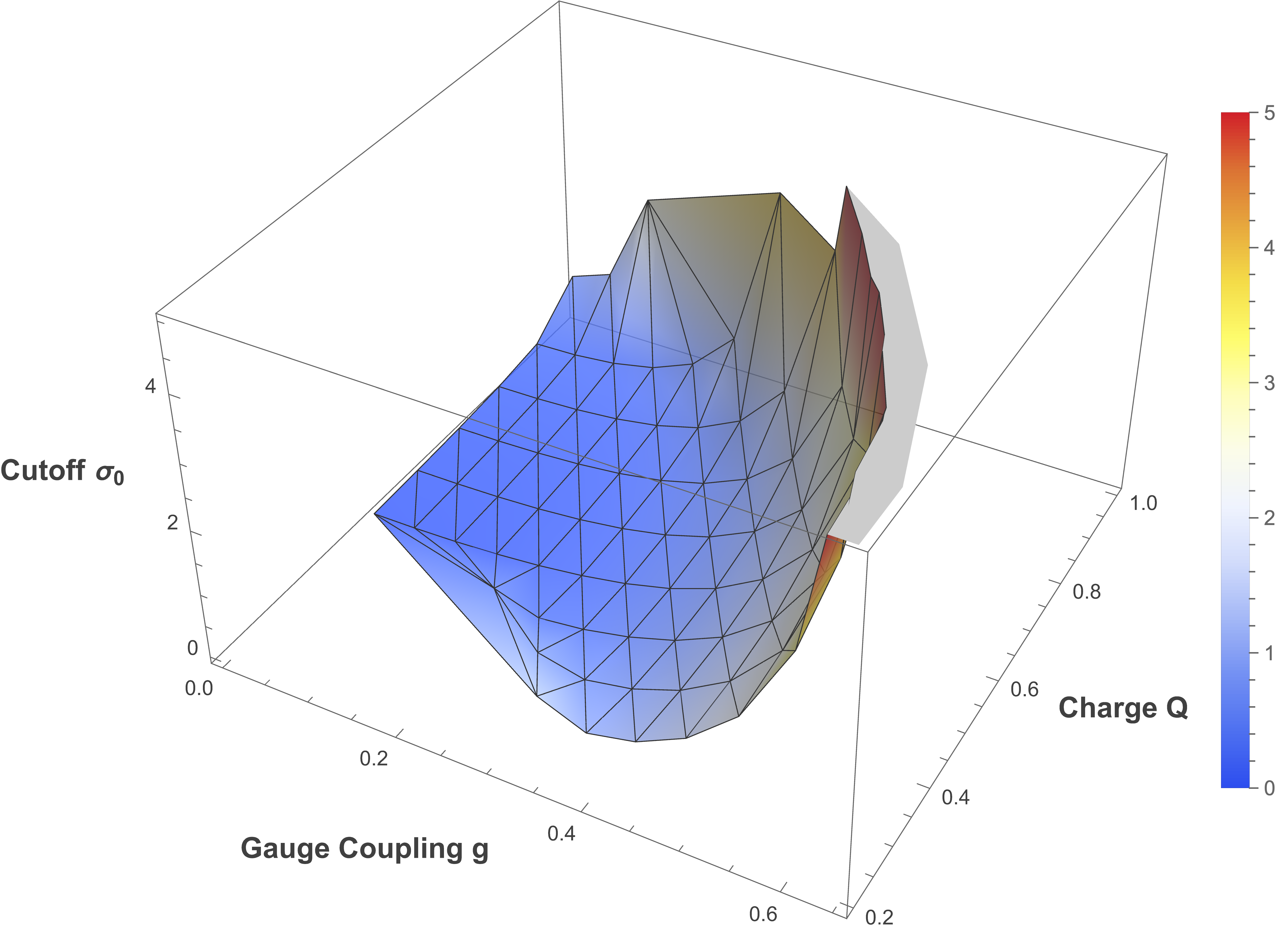}
	\caption{3D phase diagram of $\sigma_0$, showing the surface $e^{\infty}_{-} = 0$, which gives the Comisso–Asenjo activation threshold in $(g, Q)$ space. Parameters values are $a = 0.9, N_g = 0.2, v = 0.2, \xi = 0$. Regions above the surface allow energy extraction.}
	\label{fig:Sigma_PhaseDiagram_3D}
\end{figure*}

\begin{table*}%[!htbp]
	\centering
	\caption{Cutoff magnetization $\sigma_0$ for varied $Q$. Fixed: $a=0.9, N_g=0.2, v=0.2, \xi=0$.}
	\label{tab:tab1}
	\begin{tabular}{lccccccccc}
		\toprule
		& \multicolumn{9}{c}{$\sigma_0$} \\
		\cmidrule(lr){2-9}
		Coupling $g$ & $Q = 0.2$ & $Q = 0.3$ & $Q = 0.4$ & $Q = 0.5$ & $Q = 0.6$ & $Q = 0.7$ & $Q = 0.8$ & $Q = 0.9$ \\
		\midrule
		%0.000 &  &  &  & 0.004 & 0.026 & 0.059 & 0.111 & 0.213 & 0.780 \\
		%0.050 &  &  &  & 0.009 & 0.032 & 0.066 & 0.121 & 0.233 & 1.013 \\
		%0.100 &  &  &  & 0.030 & 0.051 & 0.089 & 0.154 & 0.300 & 2.451 \\
		%0.150 &  &  &  & 0.070 & 0.088 & 0.134 & 0.218 & 0.439 &  \\
		0.200 &  &  & 0.531 & 0.132 & 0.151 & 0.208 & 0.330 & 0.717 &  \\
		0.250 &  &  & 0.310 & 0.222 & 0.247 & 0.327 & 0.519 & 1.358 &  \\
		0.300 &  & 0.838 & 0.393 & 0.348 & 0.390 & 0.515 & 0.860 & 3.857 &  \\
		0.350 & 1.277 & 0.685 & 0.540 & 0.527 & 0.605 & 0.825 & 1.573 &  &  \\
		0.400 & 0.996 & 0.821 & 0.756 & 0.788 & 0.944 & 1.393 & 3.942 &  &  \\
		0.450 & 1.156 & 1.082 & 1.081 & 1.195 & 1.538 & 2.747 &  &  &  \\
		0.500 & 1.506 & 1.507 & 1.605 & 1.912 & 2.853 & 11.581 &  &  &  \\
		0.550 & 2.124 & 2.245 & 2.583 & 3.551 & 9.010 &  &  &  &  \\
		0.600 & 3.341 & 3.824 & 5.183 & 12.860 &  &  &  &  &  \\
		0.650 & 6.898 & 10.256 & 79.292 &  &  &  &  &  &  \\
		\bottomrule
	\end{tabular}
\end{table*}

In gauged supergravity geometries, the black hole spin $a$ and gauge coupling $g$ are deeply interlinked. As illustrated in Figs.~\ref{fig:Combined_Delta_Panel} and \ref{fig:properD-combined}, for near‑extremal rotation ($a=0.9$), the outer and inner horizons approach each other and eventually merge as $g$ increases, so regular black hole solutions exist only in a narrow interval of $g$. For a smaller spin ($a=0.4$), the horizons are more widely separated, and the solution remains regular for significantly larger values of $g$. Therefore, the ergoregion, and hence the possibility of energy extraction, persists over a broader range of couplings.

\begin{table}%[!htbp]
	\centering
	\caption{Cutoff $\sigma_0$ vs $Q$ (up to $Q=1.0$) for various $N_g$ ($a=0.9$, $g=0.3$, $\xi=0$).}
	\label{tab:tab2}
	\footnotesize              
	\setlength{\tabcolsep}{3pt}
	\begin{tabular}{c|ccccccc}
		\hline
		$Q$ & 0.2 & 0.3 & 0.5 & 0.6 & 0.7 & 0.9 & 1.0 \\ \hline
		0.2 &        &        &        &        &        &        &        \\
		0.3 & 0.8382 & 2.7400 &        &        &        &        &        \\
		0.4 & 0.3931 & 0.5438 & 10.8101&        &        &        &        \\
		0.5 & 0.3484 & 0.4270 & 0.9545 & 3.3253 & 0.6769 &        &        \\
		0.6 & 0.3899 & 0.4527 & 0.7407 & 1.1691 & 3.5476 &        &        \\
		0.7 & 0.5153 & 0.5774 & 0.8068 & 1.0468 & 1.6130 & 1.0421 &        \\
		0.8 & 0.8596 & 0.9252 & 1.1220 & 1.2878 & 1.5919 & 13.8087& 0.7895 \\
		0.9 & 3.8567 & 3.3378 & 2.5509 & 2.3551 & 2.3269 & 3.9601 & 19.7795\\
		1.0 &        &        &        &        & 14.4713& 4.6121 & 7.4341 \\
		\hline
	\end{tabular}
\end{table}

\begin{figure*}%[!htbp]
\centering
\includegraphics[width=0.99\linewidth]{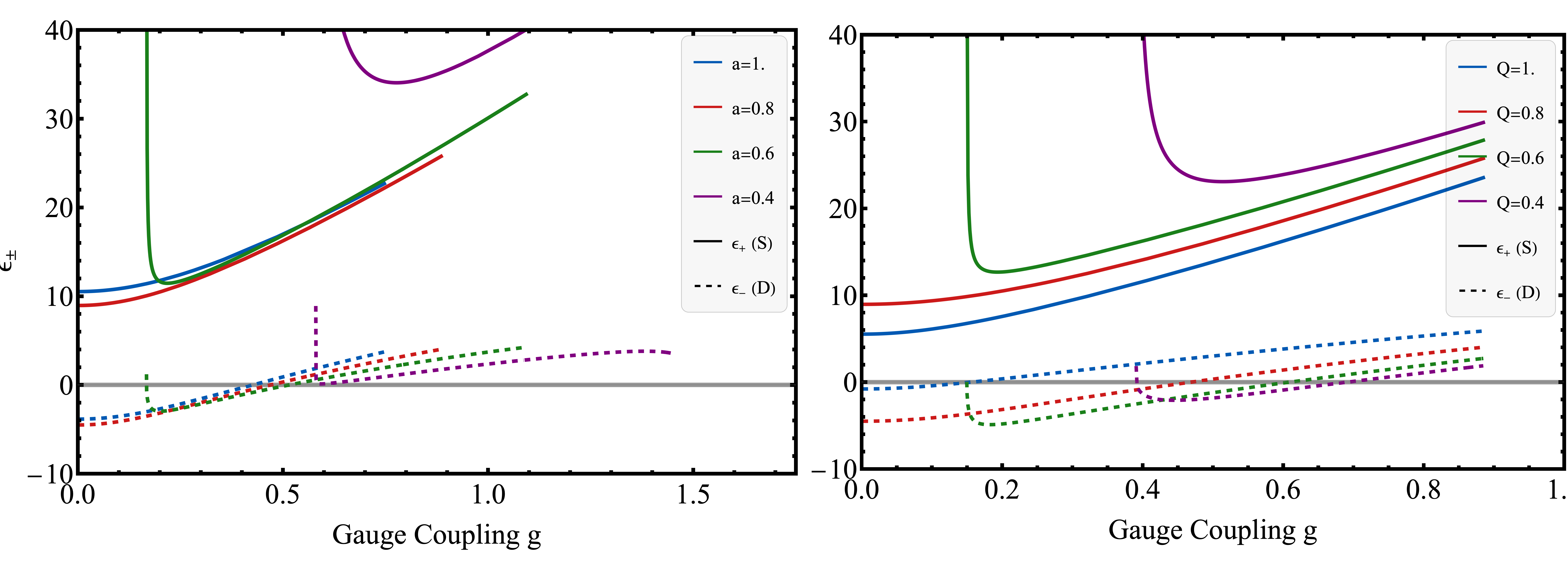}
    \caption{Asymptotic energies $\epsilon_{\pm}^{\infty}$ are shown as functions of the coupling $g$ in a rotating dyonic black hole spacetime. The solid curves denote the upper escaping branch $\epsilon_{+}^{\infty}$, while the dashed curves denote the lower negative energy branch $\epsilon_{-}^{\infty}$. The left panel illustrates the variation with spin $a$ for the fixed charge and plasma parameters. The right panel illustrates the variation with dyonic charge $Q$ for fixed spin and the same plasma parameters.}
\label{fig:Energy_vs_g-varyQa}
\end{figure*}
\begin{figure}%[!htbp]
\includegraphics[width=0.5\linewidth]{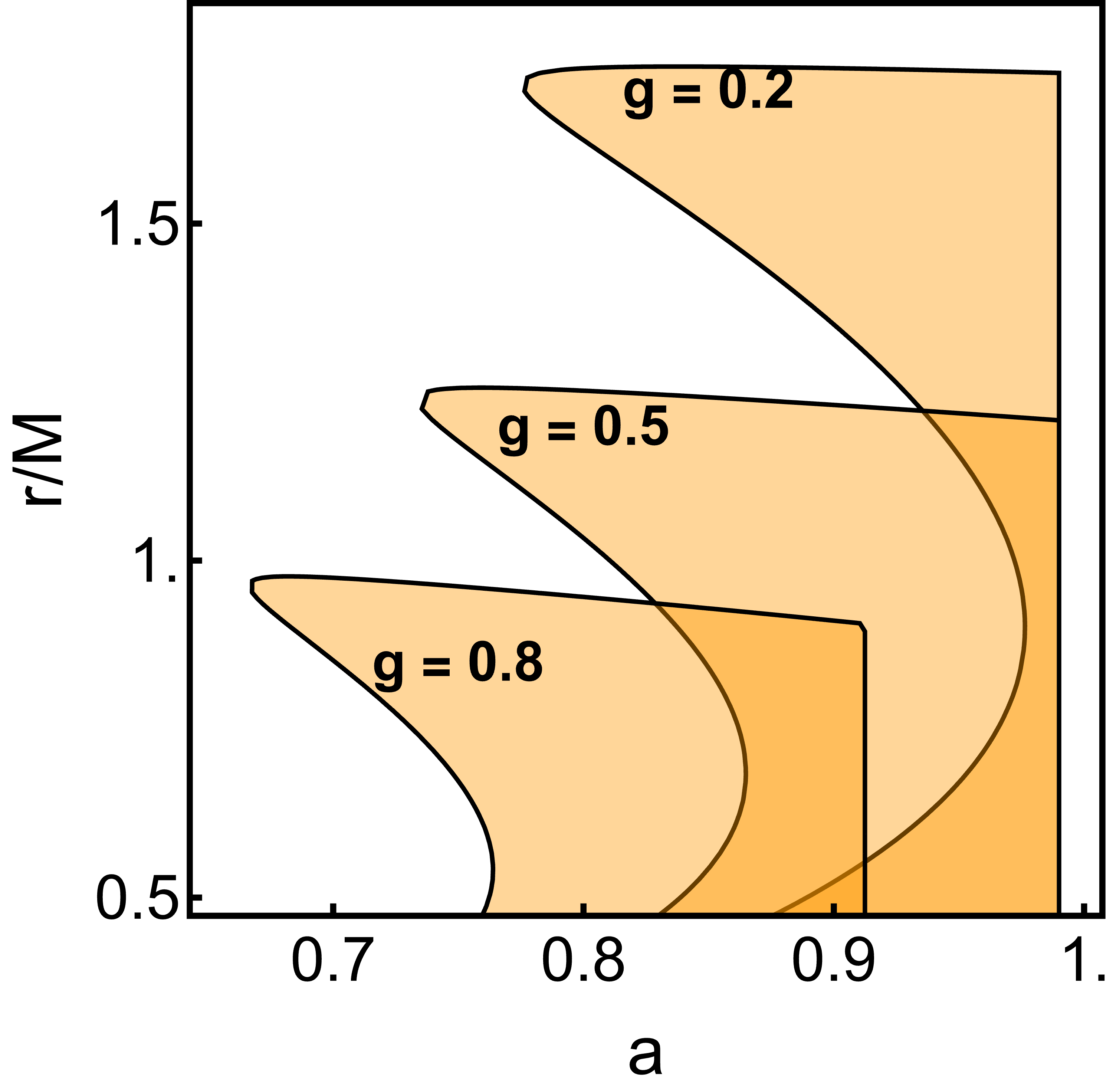}
\centering
	\caption{Regions of phase-space $(a, r/M)$ where energy extraction occurs ($\epsilon^{\infty}_{-}<0$) for different values of the gauge coupling $g\in \{0.2, 0.5, 0.8\}$ with $(Q, \xi, \sigma)=(0.1, 0, 100)$. The areas with negative $\epsilon^{\infty}_{-}$ increase as $g$ decreases, depicted in progressively lighter shades of orange.}
\label{fig:phase_space}
\end{figure}

Figure~\ref{fig:Energy_vs_g-varyQa} shows how the energies at infinity $\epsilon_\pm^{\infty}$ of the CA branches depend on the gauge coupling $g$ for the rotating dyonic $\mathcal{N}=2,\,U(1)^2$ black hole. In both panels, solid curves represent $\epsilon_+^{\infty}$ and dashed curves $\epsilon_-^{\infty}$; the horizontal black line marks $\epsilon=0$, the condition for a negative energy inflow.  In the left panel, $Q$ is fixed, and different colors correspond to spins $a=1.0, 0.8, 0.6, 0.4$. For all spins, $\epsilon_+^{\infty}$ is positive and increases with $g$. The dashed curves $\epsilon_-^{\infty}$ start negative at small $g$, increase with $g$, and cross zero at a spin-dependent critical coupling. Some branches also develop vertical divergences, where the geometry becomes extremal or the branch terminates. A higher spin shifts the zero crossing to smaller $g$ and shortens the range of couplings for which $\epsilon_-^{\infty}<0$.  In the right panel, $a$ is fixed, and different colors correspond to charges $Q=1.0, 0.8, 0.6, 0.4$. The solid $\epsilon_+^{\infty}$ curves again rise monotonically with $g$. The dashed $\epsilon_-^{\infty}$ curves behave similarly to those in the left panel; they are negative for small $g$, cross zero at a charge-dependent value of $g$, and terminate or diverge near a critical coupling. Increasing $Q$ moves both the zero crossing and the divergence to smaller \(g\), so the interval in which $\epsilon_-^{\infty}<0$ becomes narrower.  
The key takeaway is that negative energy states, which are necessary for CA energy extraction, exist only within a finite band of gauge couplings $g$. This band shrinks and shifts toward smaller $g$ as either the spin $a$ or the charge $Q$ increases, whereas the positive branch $\epsilon_+^{\infty}$ grows monotonically and never threatens the extraction condition. As a result, rapidly rotating and/or highly charged supergravity black holes have a much more restricted range of $g$ over which the Comisso–Asenjo mechanism can operate.

Figure~\ref{fig:phase_space} displays the radial extent of the region in the $(a,r/M)$ plane where Comisso–Asenjo energy extraction is possible for three gauge couplings, $g=0.2,0.5,0.8$. The horizontal axis is the spin $a$, the vertical axis is the radius $r/M$, and each shaded lobe marks the range of radii around the horizon where the negative energy branch $\epsilon_-^{\infty}<0$ exists for that $g$. The right vertical edge of each lobe corresponds to the maximal (near‑extremal) spin allowed by the coupling. From top to bottom, increasing $g$ clearly shrinks the extraction region and pushes it inward; for $g=0.2$ the lobe extends to larger radii and higher spins, whereas for $g=0.8$ it is compressed close to the horizon and terminates at a lower critical spin. Thus, the gauge coupling strongly constrains where CA extraction can operate in $(a,r)$ space; a larger $g$ reduces the radial thickness of the negative energy zone and lowers the maximum spin for which it exists. Therefore, highly gauged black holes permit energy extraction only in a small confined region near the horizon. This spatial restriction induced by the AdS confining potential stands in sharp physical contrast to Kerr-de Sitter spacetimes, where a positive cosmological constant expands the allowed parameter space and raises the maximum spin bound, thereby facilitating energy extraction further away from the horizon \cite{Wang2022}.

\subsection{Energy Extractions Rate \&  Reconnection Efficiency}

In this section, we examine the power and efficiency of energy extraction from the rotating dyonic $\mathcal{N}=2,\,U(1)^2$ black hole.  \cite{Comisso2021} proposed that these two quantities primarily depend on the rate at which plasma with negative energy-at-infinity density is absorbed. The power, denoted as $P_{extr}$, can be estimated by \cite{Comisso2021} as:

\begin{figure*}%[!htbp]
	\centering
	\includegraphics[width=0.9\linewidth]{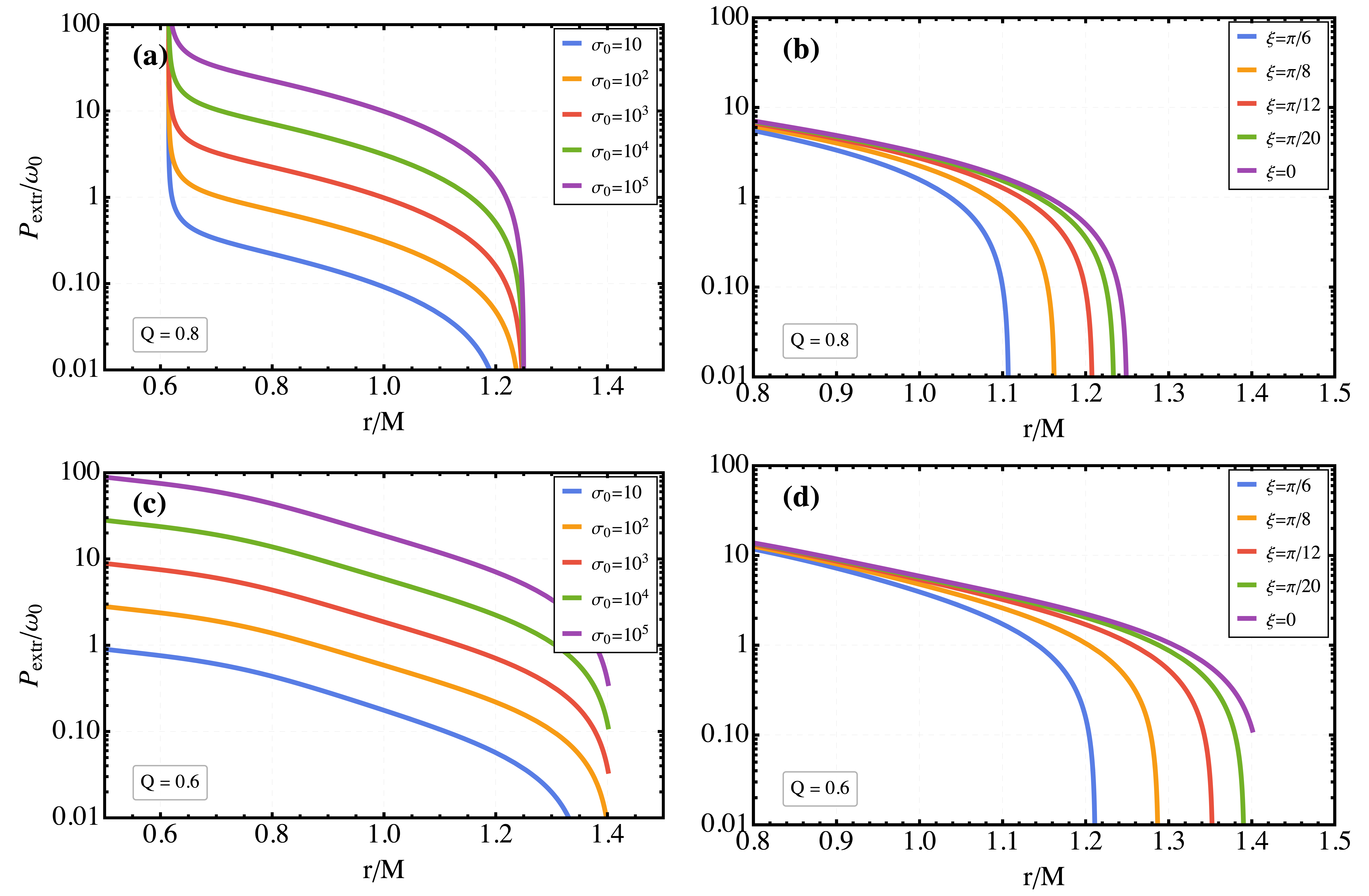}
	\caption{Extracted power per enthalpy $P_{extr}/\omega_0$ as a function of the X-point location with varying plasma magnetization (left panel) and varying orientation angle (right panel) with $(a, g, N_g, v)$ fixed at $(0.8, 0.3, 0.2, 0.2)$.}
	\label{fig:energy-per-power}
\end{figure*}
\begin{align}
	\begin{split}
		P_{extr}=-\epsilon^{\infty}_{-}\omega_0 \ A_{in} \ U_{i},
	\end{split}
\end{align}
where the reconnection inflow four-velocity $U_{in}$ is $\mathcal{O}(10^{-1})$ in the collisionless regime and $\mathcal{O}(10^{-2})$ in the collisional regime \cite{Comisso2016}. Similar power scaling and dependency on the collisionless inflow rate have been robustly demonstrated for magnetic reconnection in both regular \cite{Li2023b} and hairy \cite{Li2023a} rotating black hole spacetimes. $A_{in}$ is the cross-sectional area of the inflowing plasma, which for rapidly rotating black holes can be estimated as $A_{in} \sim r_E^2- r_{ph}^2 $, where $r_E$ and $r_{ph}$ denote the outer ergosphere radius and photon sphere radius, respectively.

We examined the efficiency of energy extraction via magnetic reconnection. This efficiency is defined by the following equation \cite{Comisso2021}

\begin{equation}\label{eq:effici}
	\eta=\frac{\epsilon^\infty_+}{\epsilon^\infty_++\epsilon^\infty_-},
\end{equation}
which shows that energy will be extracted from the rotating dyonic $\mathcal{N}=2,\,U(1)^2$ black hole only when $\eta>1$.

\begin{figure*}%[!htbp]
	\centering
	\includegraphics[width=0.99\linewidth]{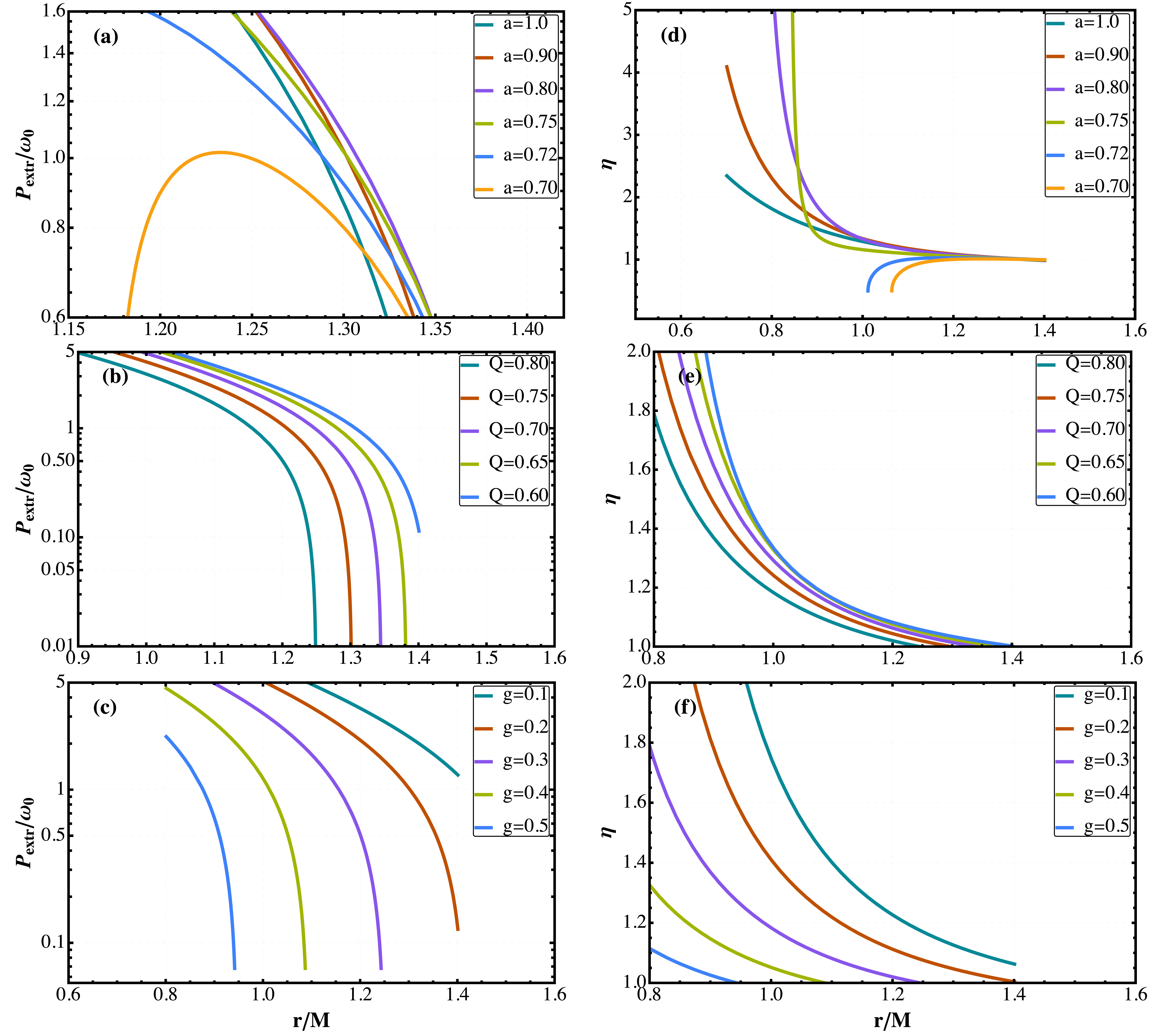}
	\caption{Extracted power per enthalpy $P_{\rm extr}/\omega_0$ (left column) and reconnection efficiency $\eta$ (right column) as functions of the X‑point radius $r/M$. 
The top row varies the spin $a$ at fixed $Q=0.8$ and $g=0.3$; 
The middle row varies the charge parameter $Q$ at fixed $a=0.8$ and $g=0.3$; the bottom row varies the gauge coupling $g$ at fixed $a=0.8$ and $Q=0.8$. 
All panels use $\xi=0$ and $\sigma_0=10^4$, and the specific parameter values are indicated in the legends.}
	\label{fig:energy-per-power_Xpt}
\end{figure*}

Figure~\ref{fig:energy-per-power} shows the energy extraction power per unit enthalpy, $P_{\rm extr}/\omega_0$ (log scale), as a function of the radius $r/M$. For $Q = 0.8$ (panels (a) and (b)), increasing the plasma magnetization $\sigma_0$ from $10$ to $10^5$ significantly increases the power at any given radius, with the curves dropping to zero near $r/M \simeq 1.25$. At the same $Q$, varying the field orientation angle $\xi$ from $\pi/6$ to $0$ shows that smaller angles yield higher power and slightly extend the maximum radius over which the extraction is possible. For $Q = 0.6$ (panels (c) and (d)), the dependence on $\sigma_0$ and $\xi$ is similar, but the lower $Q$ shifts the outer edge of the extraction region outward to $r/M \simeq 1.4$ and broadens the radial range of efficient extraction. The main takeaway is that a larger magnetization and a more azimuthally aligned field ($\xi \to 0$) lead to significantly stronger energy extraction: the power is highest close to the black hole and steadily decreases with radius, while increasing $Q$ narrows the radial range over which extraction is possible.
	
Therefore, energy extraction via magnetic reconnection is favored by lower values of the orientation angle $\xi$ and higher plasma magnetization $\sigma_0$, consistent with \cite{Zhang2024a, Ye2023}. Moreover, the extraction power strictly decreases as the X-point (the location of magnetic reconnection) moves radially outward, in agreement with \cite{Zhang2024a, Ye2023, Khodadi2023}.

Figure~\ref{fig:energy-per-power_Xpt} shows the extracted power per enthalpy, $P_{\rm extr}/\omega_0$, as a function of the X‑point location $r/M$ (left column, panels (a)– (c)) and the reconnection efficiency $\eta$ as a function of $r/M$ (right column, panels (d)– (f)) Panel (a) varies the black hole spin $a$ from 1.0 to 0.70 (see legend). All curves decrease toward zero as the X-point approaches the horizon, but their shapes differ: for most spins, the power falls roughly monotonically with decreasing $r/M$, whereas the lowest–spin case, $a=0.70$, exhibits a broad maximum before dropping to zero. Panel (d) shows the corresponding efficiencies. At larger radii, the curves for different $a$ cluster near $\eta \simeq 1$, while closer to the horizon, they separate, and some grow rapidly, indicating that for certain spins, very high efficiencies are reached only in a narrow radial range near the black hole.
Panels (b) and (e) show the variation in the charge parameter $Q$ from 0.60 to 0.80 (see legend). In both panels, increasing $Q$ lowers the curves, and in panel (b), a higher $Q$ causes the power to drop to zero closer to the horizon, shrinking the radial region where extraction is possible. Panels (c) and (f) show the variation in the gauge coupling $g$ from 0.1 to 0.5. These panels show the strongest spread: increasing $g$ sharply reduces both the extracted power and efficiency. In panel (c), for example, the curve for $g=0.5$ terminates near $r/M\simeq 0.9$, whereas for $g=0.1$ extraction remains possible up to $r/M\simeq 1.4$.

\section{Summary \& Conclusions}\label{sec:5}
Our analysis suggests two main conclusions. First, within the parameter range we considered, the cutoff magnetization $\sigma_{0,\rm cutoff}$ and even the existence of CA extraction are very sensitive to the gauge coupling $g$ and dyonic charges. Second, in this supergravity setting, high CA efficiencies can be achieved without requiring an extremal spin, and the efficiency is in fact non‑monotonic in $a$: it peaks at intermediate spin and is suppressed again as the black hole approaches its maximal rotation.

Supergravity also imposes additional geometric constraints. Our study of $\Delta_g(r)$ shows that for near‑extremal rotation ($a\simeq0.9$), the inner and outer horizons approach each other, so high‑spin solutions remain regular only in a narrow interval of $g$; outside this window, the horizon disappears, and the spacetime becomes unphysical. For lower spins (e.g., $a=0.4$), the horizons are more widely separated, the geometry is more robust against the quartic AdS term, and CA extraction remains viable up to substantially larger $g$. The NUT parameter $N_g$ and the magnetic charge further distort the ergoregion and raise $\sigma_0^{\rm cutoff}$: increasing $N_g$ makes the ergosphere more oblate, and strong topological distortion ($N_g\gtrsim1$) can be compensated only if the black hole carries a sufficiently large electric charge. In practice, only certain combinations of dyonic charges and gauge coupling admit both a regular horizon and a sizable CA‑active ergoregion.

For fixed plasma parameters ($\xi=0$, $\sigma_0=10^4$), the efficiency depends primarily on $a$, $Q$ and $g$. Decreasing $g$ while keeping $Q$ in an intermediate range enhances the extracted power and allows moderately rotating black holes (e.g., $a\simeq0.8$) to reach extraction rates and thermodynamic efficiencies that, in Kerr, would require $a\to1$. In contrast, pushing $Q$ toward extremality compresses the viable region in the radius and in $(g, Q)$ space and degrades the extraction rate. Therefore, a balance between the AdS confining scale and dyonic charges sets an upper bound on the CA power.

Unlike the Kerr case, the spin parameter in this gauged NUT‑charged geometry enters not only through the frame‑dragging term but also through the normalization factor $\Xi$ and the horizon function $\Delta_g$. Therefore, increasing $a$ modifies the redshift, shift vector, and size and location of the effective negative‑energy region. In the parameter range explored, these geometric modifications can outweigh the usual Kerr‑like enhancement from the frame dragging. Consequently, the negative-energy inflow becomes less efficient at very high spin, and both the plasma energization efficiency and the extracted power are maximized at intermediate $a$ rather than increasing monotonically with spin. This shows up in our profiles as a clear non-monotonic dependence on spin: at a fixed radius, the extracted power and efficiency are maximized at an intermediate spin ($a\sim0.8$), with both higher and lower spins yielding smaller values, and for some spins (e.g., $a=0.7$), the radial power profile itself develops a local maximum. This behavior should be interpreted as a high‑spin suppression caused by NUT/gauge deformation, not as evidence that the non‑rotating limit is intrinsically more efficient. Larger $Q$ and $g$ further quench magnetic reconnection by shrinking the CA-active portion of the ergoregion and driving the power and efficiency cutoffs inward. Across all explored parameters, the power and efficiency peak in a very thin layer just outside the horizon and then decrease over the radial range we consider, as the reconnection X‑point is pushed outward, which is in agreement with previous analyses of reconnection‑driven extraction.

On the plasma side, we found that efficient CA extraction is confined to a thin radial shell just outside the horizon and requires extreme magnetization and nearly radial outflows. Oblique pitch angles rapidly reduce both the amplitude and radial extent of power. Together with the AdS‑induced inward shift of the static limit, this makes the reconnection radius a tightly constrained quantity.

In standard Kerr spacetime, efficient CA extraction is essentially confined to almost extremal spins. In the $\mathcal{N}=2$, $U(1)^2$ supergravity background, dyonic charges and gauge coupling loosen this restriction. By raising $Q$ or lowering $g$, a moderately spinning black hole can reach power and efficiency levels comparable to those of a near‑extremal Kerr hole (cf. Figure~\ref{fig:energy-per-power_Xpt}). In this sense, gauged supergravity allows us to separately dial spin and charge and isolate the effect of each on reconnection‑driven extraction. 
This counter‑Kerr trend is consistent with \cite{Ye2023, Rudra2020}, which also found higher maximum efficiencies at lower spin. Although \cite{Rudra2020} relies on the Penrose process rather than magnetic reconnection, they demonstrated that a lower spin yields a higher maximum efficiency (e.g., $\eta_{\max}\simeq0.3988$ at $a=0.460$ versus $\eta_{\max}\simeq0.2044$ at $a=0.963$). The result of \cite{Ye2023} was obtained for a wormhole rather than a black hole, but the unique throat geometry likewise leads to higher maximum efficiencies at lower spin. Our $\mathcal{N}=2$, $U(1)^2$ dyonic gauged black hole has its own distinctive geometry and produces qualitatively similar behaviors. 
Both $Q$ and $g$ work against magnetic reconnection: increasing either parameter lowers the efficiency and total extracted power. In our plots, larger $Q$ and $g$ push the cutoff of the curves closer to the black hole, indicating that these parameters shrink the radial region of the ergosphere, where the slowed plasma can reach a negative energy state. In all panels, the highest power and efficiency consistently occurred at the smallest radii, that is, nearest to the black hole. Moving the X‑point outward strictly limits energy extraction, which is in agreement with \cite{Long2025, Ye2023, Khodadi2023}.

A natural next step is to couple this metric with concrete accretion and jet models and ask how the “breathing” ergoregion changes the ISCO, disc profile, and jet power. In particular, one can compute spectra and radio luminosities for flows in the $\mathcal{N}=2, U(1)^2$ background and compare them with Kerr‑based fits to determine whether variations in $Q$, $N_g$, and $g$ leave detectable signatures.
From an observational point of view, the non-monotonic spin dependence we find could turn reconnection-driven extraction into a probe of departures from Kerr, at least within the parameter range we have explored. In GR, most extraction scenarios predict that the jet and coronal efficiencies rise monotonically with spin; therefore, very high efficiencies are usually taken as circumstantial evidence for near‑extremal Kerr black holes. In our $\mathcal{N}=2$, $U(1)^2$ gauged supergravity background, however, the efficiencies can instead peak at moderate spin once the gauge coupling and dyonic charges are considered. In this case, a system with only modest spin could still power a highly efficient jet, whereas a nominally similar high-spin source might appear comparatively underluminous. If future joint measurements of spin (e.g., from X‑ray spectroscopy) and jet or coronal energetics (radio and X‑ray) reveal systematic deviations from the Kerr trend that higher spin corresponds to higher efficiency, such trends could point to either non‑Kerr geometry or additional matter fields of the type realized in string‑motivated supergravity models.

\subsection*{Acknowledgments}
S.R. and L.R. acknowledge the support from the Grinnell College CSFS's grants program and extend gratitude to the Center for Astrophysics $|$ Harvard \& Smithsonian for their hospitality during the completion of this work.

\subsection*{DATA AVAILABILITY} 
This study was purely theoretical. All results were derived analytically from the equations presented in the main text and were reproducible using the methods and parameters specified. No empirical data or numerical simulations requiring archival storage were used.

\end{justify}

\end{document}